\documentclass{aa}  
\usepackage{graphicx}
\usepackage{txfonts}
\usepackage{comment}
\usepackage{natbib}

\renewcommand{\vec}[1]{\mathbf{#1}}                                                                                     % vector symbol
\newcommand{\earth}{\text{earth}}                                                                                                       % earth
\renewcommand{\sun}{\text{sun}}                                                                                                         % sun
\newcommand{\R}{{\mathbb R}}                                                                                                                    % real numbers
\newcommand*{\norm}[1]{\left\lVert#1\right\rVert}                               % norm

\usepackage{ulem}

\usepackage{color}

\begin{document}

   \title{Shape and spin distributions of asteroid populations from brightness variation estimates and large databases}
        \titlerunning{Shape and spin distributions of asteroid populations}

   %\subtitle{}

   \author{H. Nortunen\inst{1}, M. Kaasalainen\inst{1},                                 
          J. \v{D}urech\inst{2},
          H. Cibulkov\'a\inst{2},
          V. Ali-Lagoa\inst{3,4}, and 
J. Hanu\v{s}\inst{2}}

   \institute{Tampere University of Technology, Department of Mathematics, 
PO Box 553, 33101 Tampere, Finland              
         \and
             Charles University, Faculty of Mathematics and Physics, Astronomical Institute, V Holešovickách 2, 18000 Praha 8, Czech Republic
                                                \and
                                                Observatoire de la C\^{o}te d’Azur, Boulevard de l’Observatoire, CS 34229, 06304 Nice cedex 4, France
                                                \and
                                                Max-Planck-Institut f{\"u}r extraterrestrische Physik, Giessenbachstrasse 1, 85748 Garching, Germany}

   \date{Received February 21, 2017; accepted March 01, 2017}

\authorrunning{H. Nortunen et al.}

% \abstract{}{}{}{}{} 
% 5 {} token are mandatory
 
  \abstract
   {Many databases on asteroid brightnesses (e.g. ALCDEF, WISE) are potential sources for extensive asteroid shape and spin modelling. Individual lightcurve inversion models require several apparitions and hundreds of data points per target. However, we can analyse the coarse shape and spin distributions over populations of at least thousands of targets even if there are only a few points and one apparition per asteroid. This is done by examining the distribution of the brightness variations observed within the chosen population.}
   {Brightness variation has been proposed as a population-scale rather than individual-target observable in two studies so far. We aim to examine this approach rigorously to establish its theoretical validity, degree of ill-posedness, and practical applicability.}
   {We model the observed brightness variation of a target population by considering its cumulative distribution function (CDF) caused by the joint distribution function of two fundamental shape and spin indicators. These are the shape elongation and the spin latitude of a simple ellipsoidal model. The main advantage of the model is that we can derive analytical basis functions that yield the observed CDF as a function of the shape and spin distribution. The inverse problem can be treated linearly. Even though the inaccuracy of the model is considerable, databases of thousands of targets should yield some information on the distribution. We employ numerical simulations to establish this and analyse photometric databases that provide sufficiently large numbers of data points for reliable brightness variation estimates.}
   {We establish the theoretical soundness and the typical accuracy limits of the approach both analytically and numerically. We propose a robust brightness variation observable $\eta$ based on at least five brightness points per target. We also discuss the weaker reliability and information content of the case of only two points per object. Using simulations, we derive a practical estimate of the model distribution in the (shape, spin)-plane. We show that databases such as Wide-field Infrared Survey Explorer (WISE) yield coarse but robust estimates of this distribution, and as an example compare various asteroid families with each other. }
{}
 % {The observable $\eta$ is applicable for obtaining information about shape and spin properties of asteroid populations. It is effective for %discovering the existence of statistical peaks, as well as the approximate location of the peaks. The method is robust, as long as there are enough %observed asteroids available in the data source. Fortunately, such databases exist and are accessible.}


   \maketitle
%
%________________________________________________________________

\section{Introduction}

Most of the current roughly one thousand asteroid shape and spin models, such as the ones given in the Database of Asteroid Models from Inversion Techniques (DAMIT\footnote{http://astro.troja.mff.cuni.cz/projects/asteroids3D/web.php.}), are based on photometry \citep{genproj,aiv}. Databases from large sky surveys and the inversion methods of sparse lightcurves \citep{sparse,sparseD,wise} will greatly expand the list of models of individual asteroids. However, the databases also contain measurements that are not sufficient for individual models, but nevertheless can be expected to provide information on the statistical shape and spin distributions of the observed asteroid populations. Such measurements are, for example, brightness sequences ranging from a few points to full lightcurves. These can be transformed into statistical data by examining the population-level distribution of the brightness variation within each observed sequence.

The variation among each target's brightnesses, sampled over a wide range of rotation phases, can mostly be attributed to the shape elongation and the sub-Earth aspect angle of the object. The more detailed shape of the body, and especially its irregularity, is another important factor, but this cannot be included in a population-level model due to its complexity. The elongation and aspect have simply describable effects on the brightness variation, with monotonous dependencies. The detailed illumination and viewing geometry also have a somewhat complicated effect. Fortunately, if we include these factors as a part of the modelling error by simply using opposition geometry in the model, they will not make a large contribution to the total error budget as we will discuss below.

A realization of the statistical approach was presented by \citet{szabo}. They included over $10^4$ pieces of pairwise brightness differences and, while their study did not contain analytical or numerical inspection of the generic inverse problem, they concluded that 
a statistical analysis is possible. The asteroid populations were characterized by shape elongation distributions. A similar type of observable and method was used by \citet{mcneill}. We aim to establish the usefulness of the statistical approach by investigating the inverse problem both analytically and numerically by simulations, including the role of the insufficient model and other assumptions that do not necessarily hold in practice. We also seek to define a good observable of the brightness variation such that its information content is as high as possible. 

We generate cumulative distribution functions (CDFs) of the brightness variation levels observed within large asteroid populations, aiming to study what the CDF reveals about the properties of the population. We choose the CDF since it is the most direct, well-defined, and stable data product describing the distribution statistics of a one-dimensional observable. Morever, the analytical study of the inverse problem requires the CDF integral of the model in the first place. To keep our model CDF simple and solvable, we choose to utilize as few parameters as possible, namely the shape elongation $p$ and the spin $\beta$. We define $p$ as the ratio of the equatorial widths of the asteroid, and $\beta$ is the ecliptic polar angle of the spin axis. The main principle is to derive analytical basis functions that describe the contribution of the proportion of targets in a given $\beta$ and $p$-bin to the observed CDF. These functions allow both the inspection of the information content of the data and the use of robust inversion methods. In particular, we show that  it is possible to obtain information about the $\beta$ distribution in addition to $p$.

In addition to the thorough analysis of the inverse problem, one of our main goals is to introduce an especially useful observable, $\eta$, that is a measure of the variation of the squared intensities of a sequence. The estimate $\eta$ can be employed in a variety of contexts. \citet{cib} used $\eta$ to investigate brightness data that were not sufficient for sparse lightcurve inversion but suitable for creating a number of most probable simple asteroid models. These were used especially to demonstrate the slight periodic anisotropy of the distribution of rotation longitudes.

As the shape and spin distributions of asteroid populations are complex to interpret by themselves, we aim to introduce a tool for comparing the distributions of different populations. We do not make astronomical interpretations of the populations or their differences, but our objective is to show that the CDF-based method is a useful tool for the statistical investigation of populations. 

We use asteroid databases that provide a number (usually at least five) of points for effectively one rotation: that is, obtained essentially randomly within a few nights such that the aspect angles of the Earth and the Sun are effectively constant. This allows the use of analytical basis functions for the CDFs as well as a rigorous study of the inverse problem. Other scenarios can be used as well, but these require additional assumptions and/or purely numerical treatment, further increasing the model noise.

In Sect.\ 2 we formulate the observables and the forward problem of the derivation of a CDF from the population model. In Sect.\ 3 we discuss the solution methods of the inverse problem and prove its fundamental uniqueness and stability properties. In order to verify the applicability of our method and obtain information on the level of error, we perform realistic simulations in Sect.\ 4 to assess the information content in practice.
In Sect.\ 5 we use observations from databases to analyse asteroid families. We do not consider observational biases here: we simply take the available data at face value and analyse them as such. Additionally, we introduce a tool for a statistical comparison of distinct families. Conclusions are presented in Sect.\ 6, and mathematical details are given in three appendices.

\section{Observables and forward problem}

In this section, we define our model and formulate the forward problem. We introduce our main observable, denoted as $\eta$, based on the variation of squared brightness intensities, and show why it is useful both analytically and by its information content. We also briefly consider the applicability of measurements with only two observed points. Mathematical details are given in Appendices \ref{app:basis} and \ref{app:2point}.

Our model shape is the triaxial ellipsoid, since it has a particularly simple analytical expression for the area of its projection in any given viewing direction \citep{con}. In this paper, we use the terms brightness and projection area interchangeably, because they are physically almost the same (up to a scaling factor) for dark targets when the viewing and illumination directions coincide \citep{genproj}. We further simplify the model
(semiaxes $a,b,c$) with $b=c=1$ and use $p:=b/a$ for
describing the shape elongation (the smaller the $p$, the more elongated the body).  This is a coarse
shape approximation for individual targets of general shape, but even if our model is actually not very realistic in practice,
it should portray some coarse-scale population tendencies correctly when we have many
observations. Indeed, as we will show by simulations, it suffices to have a model that represents
the effects of shape elongation and spin direction in a roughly correct manner.

\subsection{Amplitude $A$ and its CDF $C(A)$}

Let the polar aspect angle of the viewer be given by $\theta$: $\cos\theta={\mathbf v}\cdot{\mathbf e}$, where $\mathbf v$ is the spin direction (given by the polar coordinates $(\beta,\lambda)$ in the inertial frame) and $\mathbf e$ the line of sight (unit vectors). Due to model symmetry, we only need to consider the interval $0\le\theta\le\pi/2$.
With $\phi$ for the longitudinal angle in a coordinate frame fixed to the ellipsoid, the area $I$ of the ellipsoid's projection in the direction $\mathbf e$ is \citep{con}
$$
I=\pi abc\sqrt{\frac{\sin^2\theta\cos^2\phi}{a^2}+\frac{\sin^2\theta\sin^2\phi}{b^2}+\frac{\cos^2\theta}{c^2}}.
$$
In terms of our model definitions, the brightness $L$ scaled against the maximal possible value $\pi a$ is
\begin{equation}
\begin{split}
L&=\sqrt{p^2\sin^2\theta\cos^2\phi+\sin^2\theta\sin^2\phi+\cos^2\theta}\\
&=\sqrt{1+(p^2-1)\sin^2\theta\cos^2\phi}.\label{bright}
\end{split}
\end{equation}

The statistical observable can be anything that describes the variation of the brightness as the target rotates (at a fixed $\theta$). A simple version is the peak-to-peak amplitude; here we consider the ratio $A=L_{\rm min}/L_{\rm max}=L\vert_{\phi=0}/L\vert_{\phi=\pi/2}$ (i.e. an "inverse amplitude": the smaller the $A$,
the larger the variation). Thus we have chosen the convenient $0< p\le 1$ and $0<A\le 1$ (rather than either of these extending to infinity). We would like to note that the amplitude $A$ is based on intensity; we do not use magnitudes anywhere. The
assumption is that all objects rotate about an axis (the ellipsoid's $c$-axis), which produces the observed projections random in $\phi$. At first, we consider the randomness of $\theta$ to be due to the uniform distribution of rotation axis directions on the unit sphere $S^2$; later, we take the $\theta$-distribution to be caused by a shifting viewing position.

The
amplitude $A$ is given by
\begin{equation}
A=\sqrt{\cos^2\theta+p^2\sin^2\theta}=\sqrt{1+(p^2-1)\sin^2\theta}.\label{amp}
\end{equation}
Using the amplitude, we can derive analytical basis functions, the linear combination of which yields the CDF $C(A)$ of a population with a given distribution of $p$ (and $\beta$); see Appendix \ref{app:basis} for details.

\begin{figure}
\centering \includegraphics[width=0.5\textwidth]{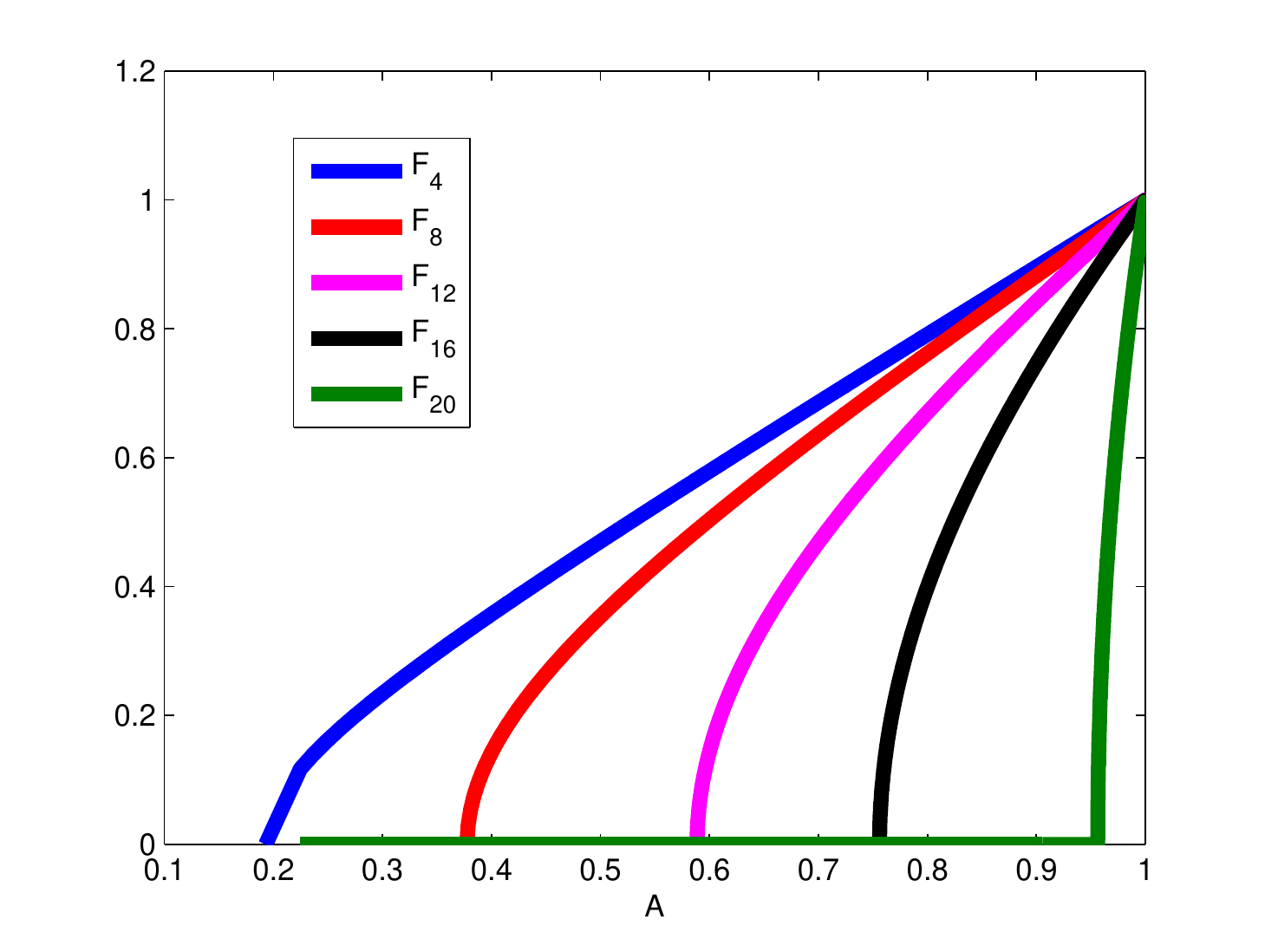}
\caption{Sample basis functions $F_i$ on a set of bins $p_i$, where $i = 1$, $\ldots$, $20$.}
\label{Fi}
\end{figure}

In an approximation consistent with the coarseness of the model, it is practical to divide the population under study into a moderate number $n$ of bins in each of which all members have the same $p$ (and $\beta$). Then, if we have only
$p$-bins and isotropic $\theta$, the CDF of the values of $A$ observed in the model population is
\begin{equation}
C(A)=\sum_{i =1}^n w_i\,F_i(A),
\label{eq:lineq00}
\end{equation}
where the basis functions $F_i(A)$ are, from Eq. (\ref{CAp}),
\begin{equation}
F_i(A)=\left\{\begin{array}{rl}
0, & A\le p_i\\
\sqrt{\frac{A^2-p_i^2}{1-p_i^2}} ,& A>p_i.
\end{array}\right.
\label{Fieq}
\end{equation}
The range of the monotonously increasing $F_i$ is $[0,1]$, and $F_i=1$ at $A=1$ (Fig.\ \ref{Fi}). The occupation numbers of the bins are given
by $w_i$.

Let us now include the $\beta$-distribution\footnote{We note that our $\beta$ is measured from the pole: $0\le\beta\le\pi$. The ellipsoidal model, however, folds here all solutions of $\beta$ into the interval $0\le\beta\le\pi/2$; that is, the model cannot distinguish between pole latitudes above and below the ecliptic plane.}
by assuming that there is a concentration of viewing geometries towards the ecliptic plane (see Appendix A). If we assume a $(p_i,\beta_j)$-grid, $i=1$, $\ldots$, $l$ and $j=1$, $\ldots$, $m$, then we have $n = lm$ bins in the grid. We can write the CDF $C(A)$ as
\begin{equation}
C(A)=\sum_{ij} w_{ij}\,F_{ij}(A),
\label{eq:lineq01}
\end{equation}
where, from Eq.\ (\ref{CApx}), the monotonously increasing basis functions $F_{ij}(A)$ with the range $[0,\pi/2]$ are,
\begin{equation}
F_{ij}(A)=\left\{\begin{array}{rl}
0, & A\le p_i\\
\frac{\pi}{2}-\arccos\frac{\sqrt{A^2-p_i^2}}{\sin\beta_j\sqrt{1-p_i^2}} ,& p_i<A<\mathcal{F}(p_i, \beta_j)\\
\frac{\pi}{2}, & A\ge\mathcal{F}(p_i, \beta_j),
\end{array}\right.\label{Fijeq}
\end{equation}
where $\mathcal{F}(p_i, \beta_j) = \sqrt{\sin^2\beta_j+p_i^2\cos^2\beta_j}$.
The $F_{ij}(A)$ are sigmoidal functions (Fig.\ \ref{Fij}), approaching the step function when $p_i\rightarrow 1$ (step at $A=1$) or $\beta_j\rightarrow 0$ (step at $A=p_i$). Because of our choice of scale of $p$ and $A$, parts of the $F_{ij}$ tend to pack together at the low end of $A$, making them less well distinguishable than those with the slope in the higher end of $A$, but on the other hand, $p$-values less than 0.4 are not likely for real celestial bodies.

The occupation numbers $w_{ij}$ are assigned to each bin. It should be noted that occupation levels proportional to $\sin\beta$ mean a uniform density on the direction sphere: that is, a constant $f(\beta)$. For applications, we adopt the convention of reporting the actual (relative) target numbers $w_{ij}$ for a given $\beta$-slot (absorbing the factor $\sin\beta$), and we plot these as the density functions (DF; number densities in $\beta$ rather than on the sphere) in the following sections.

\subsection{Brightness variation $\eta$} \label{sec:eta}

If the amplitude cannot be measured directly, a practical observable is the brightness variation around some
mean value, requiring fewer points. Using intensity squared, $L^2$, to get rid of the square root in integrands,
we obtain from Eq.\ \eqref{bright} a simple average quantity over model rotation
at a constant $\theta$:
$$
\langle L^2\rangle
=\frac{1}{2\pi}\int_0^{2\pi} \Big( 1+\sin^2\theta(p^2-1)\cos^2\phi \Big) \,d\phi=
1+\frac{1}{2}\sin^2\theta(p^2-1).
$$
Now, a measure of variation\footnote{Other definitions could be used as well, but this form leads to simple closed-form formulae.} for $L^2$ over a rotation is
\[
\begin{split}
\Delta (L^2)&=\sqrt{\langle(L^2-\langle L^2\rangle)^2\rangle}=
\sqrt{\langle[\sin^2\theta(p^2-1)(\cos^2\phi-1/2
%\frac{1}{2}
)]^2\rangle}\\
&=\sin^2\theta(1-p^2) \left[ \frac{1}{2\pi}\int_0^{2\pi}(\cos^4\phi-\cos^2\phi)\,d\phi
+\frac{1}{4} \right]^{1/2}\\
&=\sin^2\theta(1-p^2)/\sqrt{8},
\end{split}
\]
and normalizing this with $\langle L^2\rangle$ yields
\begin{equation}
\begin{split}
\eta(\theta,p)&:=\Delta (L^2)/\langle L^2\rangle
=\sqrt{ \Big\langle \Big( \frac{L^2}{\langle L^2\rangle}-1 \Big)^2 \Big\rangle }\\
&=\frac{1}{2\sqrt{2}}\Big[\frac{1}{\sin^2\theta(1-p^2)}-\frac{1}{2}\Big]^{-1}.
\end{split}\label{eta}
\end{equation}
We note that $0\le\eta\le1/\sqrt{2}$.
Thus, by Eq.\ (\ref{amp}), our brightness variation $\eta$ is directly related to the amplitude $A$:
\begin{equation}
\eta=\frac{1}{\sqrt{8}}\Big(\frac{1}{1-A^2}-\frac{1}{2}\Big)^{-1},
\quad A=\sqrt{1-\Big(\frac{1}{\sqrt{8}\eta}+\frac{1}{2}\Big)^{-1}}.\label{eq:eta-A}
\end{equation}
This is a particular advantage of the biaxial model: we can use all available estimates of $A$ (available for dense lightcurves) and $\eta$ together to form a $C(A)$. One can also directly compute a $C_\eta(\eta)$ with a procedure similar to that of Appendix A, resulting in similar types of integrals, but we choose the $A$-based formulation as it is more intuitive and leads to simpler equations. The end result is naturally the same in both cases. For the triaxial ellipsoid, a similar simple conversion between $A$ and $\eta$ is not possible since $\eta$ would depend on $\theta$ (and $b$) in addition to $A$ \citep{cib}.

The condition $0 \le \eta \le 1/\sqrt{2}$ from Eq.\ \eqref{eta} may be violated at some measurements of $L(\phi, \theta, p)$ when the parameter $p$ is low ($\lesssim 0.4$). The maximal theoretical value of $\eta$ for all lightcurve shapes (not just those from ellipsoids) approaches one, given by a boxcar-shaped lightcurve with half of the values at a constant level and half approaching zero. Real lightcurves have lower values of $\eta$ because the lightcurve is smoother than the step-function type. If $\eta > 1/\sqrt{2}$ (this may happen due to an irregular shape, outliers, and/or particular spacing of the sample points), it follows that the amplitude $A$ becomes purely imaginary according to Eq.\ \eqref{eq:eta-A}. For computational purposes, we have omitted complex amplitudes in our study (these are rarely encountered). %In practice, the majority of realistic values of $p$ are in the range of $[1/2, 1]$, so it is rare that complex amplitudes are encountered.

\subsection{Two-point brightness variation} \label{sec:2point}

The accuracy of the $\eta$ estimate depends on the number of data points (and their coverage of the rotational phase)
used to approximate  $\Delta L^2/\langle L^2\rangle$. To analyse the information content of the minimal case of two points per rotation, we briefly consider simple
pairwise brightness differences. For a group of $N$ points for one target, the number of such values is 
$N(N-1)/2$, ordered such that the difference $0<q\le 1$ is
$q=L_{\rm dimmer}/L_{\rm brighter}$. We do not need to have more than one such pair for
one target, so one object does not have to cover
the rotational phases well. This is the observable used in \citet{szabo}. We examine its properties from the inversion point of view in Appendix \ref{app:2point}. Since they turn out to be considerably inferior to those of $\eta$ (above all, no information can be obtained on the distribution of $\beta$), we do not consider the two-point data further in the main text or database analysis.

\citet{mcneill} used a similar type of observable, with the two points connected by a short time interval (effectively yielding the slope of a lightcurve). This problem is even more complicated as it necessarily introduces the rotation period of the target into the forward model, with overlapping effects of $p$, spin, and period distributions (see Appendix \ref{app:2point}). Thus, in practice, this observable necessitates the heavy use of a priori assumptions in the inverse problem. Again, this is outside our aim of minimal use of parameters and prior functions, so we do not consider the slope version of two-point data further.

 \section{Inverse problem}

In this section, we consider the fundamental properties of the inverse problem version of the forward model above 
before moving to realistic shapes and numerical results in the following sections. In particular, we present and prove a uniqueness result that shows how the distributions of both $p$ and $\beta$ can be uniquely obtained from the CDF $C(A)$. That is, we show why $\eta$-data contain unambiguous information on $f(p,\beta)$. This may seem counterintuitive at first glance, since the effects of $p$ and $\beta$ are certainly mixed for a single observation of $\eta$ (i.e. a lightcurve).
The point is that, under the ecliptic-plane assumption of Appendix A, the distribution of $\eta$ in a large population separates the effects from each other.

For the $p$-only case of isotropic $\theta$, the inverse problem can be cast linearly in matrix form. From Eq.\ \eqref{eq:lineq00}, we write
\begin{equation}
Mw = C ,
\label{eq:lineq02}
\end{equation}
where $C\in\R^k, w\in\R^n,\, M_{ji}=F_i(A_j)$, and $F_i(A_j)$ are given by Eq.\ \eqref{Fieq}.
The $k$ observed values of $A$ (derived from $\eta$) are sorted in ascending order, and the vector $C$ contains the observed CDF: each element $C_j = j/k$ is the value of $C(A_j)$. In Appendix \ref{app:stability}, we discuss the analytical stability results of the distribution function $f(p)$ obtained from $C(A)$ (i.e. $\eta$-scatter data). In particular, we show that the inverse problem is not strongly ill-posed: the errors in the details of the observed CDF do not amplify fast in the error of the recovered $f(p)$.

In the inverse problem of full $f(p,\beta)$, we can write Eq.\ \eqref{eq:lineq01} in the form of Eq.\ \eqref{eq:lineq02} as well; now
$C\in\R^k$, $w\in\R^n$ and $M$ is a $k \times n$-matrix ($n = lm$),
$$
M = \begin{pmatrix} F_{11}(A) & \ldots & F_{1m}(A) & \ldots & F_{l1}(A) & \ldots & F_{lm}(A)  \end{pmatrix} ,
$$
and the occupation numbers $w_{ij}$ are given in $w$ with indexing similar to $M$ above.

\begin{figure}
\centering \includegraphics[width=0.5\textwidth]{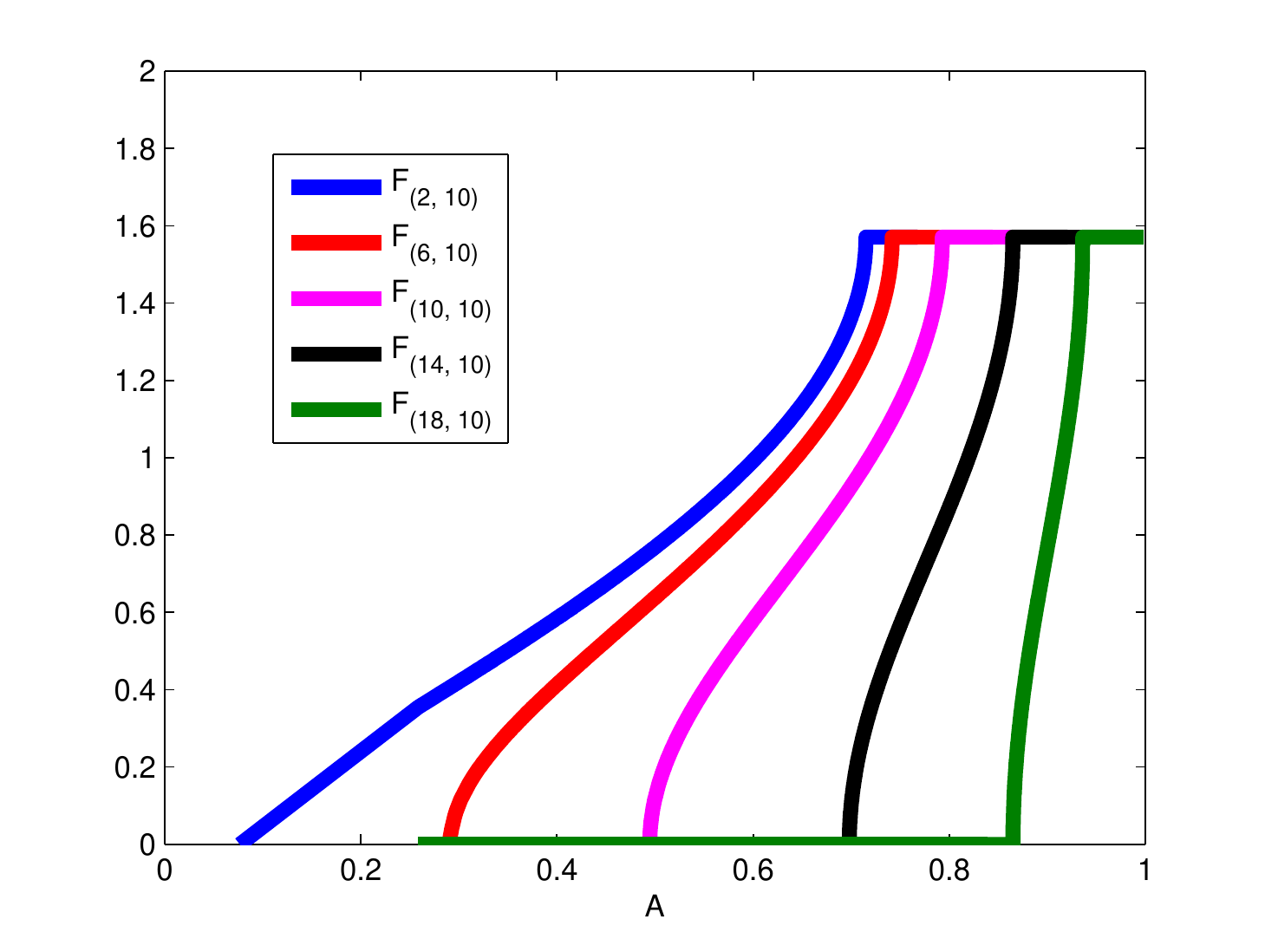}
\caption{Sample basis functions $F_{ij}$ on a set of bins $(p_i , \beta_j)$, where $i = 1$, $\ldots$, $20$ and $j = 1$, $\ldots$, $19$. The shape of the basis functions shows that they are linearly independent.}
\label{Fij}
\end{figure}

\textbf{Uniqueness result.} Perhaps surprisingly, the distribution of both $p$ and $\beta$ is unambiguously recoverable; that is, the bin model coefficients $w_{ij}$ are uniquely determined by the $C(A)$. To show this, we notice that the pairs of end points (i.e. the values of $A$ between which $F_{ij}$ changes: $A_-$ at $F_{ij}=0$ and $A_+$ at $F_{ij}=\pi/2$), are unique for each $F_{ij}$. Any combination of $F_{ij}$ will start to deviate from zero at the lowest $A_-$ of the set, and stop changing at the highest $A_+$ of the set. Thus both end points of an $F_{ij}$ cannot be matched by a superposition of other $F_{rs}$, so the $F_{ij}$ are linearly independent (see Fig.\ \ref{Fij} for illustration). Since the model $C(A)$ is a linear combination of the $F_{ij}$, the $w_{ij}$ are unique for the observed $C(A)$. As $w_{ij}$ are the occupation numbers of each $(p_i, \beta_j)$ bin, \textbf{this proves that the full $p$ and $\beta$ distribution is uniquely obtained for the ecliptic-orbit model}.

If we want to use regularization to smooth the solutions for either $p$ or $\beta$, we may apply, for example, the following $(n-1)\times n$ regularization matrix in the $p$-only case:
$$
(R_p)_{ij} = \left\{ \begin{array}{rl}
-1/(p_{i+1}-p_i), & i=j \\
1/(p_{i+1} -p_i), & j=i+1 \\
0, & {\rm elsewhere}
\end{array} \right.
$$
and its generalization for the $(p,\beta)$-grid, as well as similarly $R_\beta$ with $\beta$. These approximate the gradients at each $w_{ij}$ in the $p$- and $\beta$-directions only; one can construct more general matrices, but we found these to suffice for our problem. The occupation numbers can be obtained as a solution to an optimization problem:
\begin{equation}
\hat w = \arg\min_w \left( \Vert{C-Mw}\Vert^2 + \delta_p \Vert{R_p w}\Vert^2 + \delta_{\beta} \Vert{R_{\beta} w}\Vert^2 \right),\quad w\in\R^n_+ .\label{opt}
\end{equation}
To obtain the solution $\hat{w}$, we create an extended matrix $\tilde{M}$:
\begin{equation}
\tilde M=\left(\begin{array}{r}
M\\
\sqrt\delta_p R_p\\
\sqrt\delta_{\beta} R_{\beta} \end{array}\right), \quad \tilde C=\left(\begin{array}{l}C\\0_{(l-1)m}\\
0_{l(m-1)}\end{array}\right),\label{extMC}
\end{equation}
assuming a $(p, \beta)$-grid of the size $n=lm$ with, respectively, $l$ and $m$ equally spaced $p$- and $\beta$-values, and we find the least-squares solution of $\tilde{M} w = \tilde{C}$ with the constraint that each element of $w$ be larger than or equal to zero. The extended vector $\tilde{C}$ due to the regularization ensures that there are always more equations than unknowns regardless of the number of original data points in $C$. Due to the instability of the problem, the direct unconstrained matrix solution would lead to negative values, but in for example the Matlab environment, the positivity constraint is simple to enforce with a standard function. We found that this is more practical than nonlinear optimization with, for example, $w_i = \exp(z_i)$. 

We emphasize here that, despite the similar fitting procedures, CDFs are quite different from lightcurves as data. First of all, noise does not show as signal deviations because CDFs are monotone curves.
The error in the observed CDF curve is essentially due to convolution (the distribution function is multiplied by the error probability function under the CDF integral), causing the smoothing of the curve (very noisy data would produce a featureless CDF resembling a step function).
In our analysis, we do not attempt to deconvolve the original CDF, since the convolution (i.e. error) function is not known: in addition to the random and systematic brightness errors, it depends on the number and temporal distribution of the data points. Also, the addition of more measurements does not fill the gaps between points as in lightcurves: it alters the whole shape of the CDF. A visually good density of points in a CDF does not imply that its shape is near the correct theoretical one from infinitely many points. Computationally, one can use very high densities for interpolating between the actual CDF values to construct the values in the data vector. Then one does not have to use the full high number of observations, which may be helpful for software dealing with the positivity constraint of the solution based on Eq.\ \eqref{extMC}.

Obviously $p$-values lower than about 0.4 start to become unrealistic, so one could also use an additional regularization function and a lower limit on $p$-values. We have, however, used the whole scale since the ostensibly unrealistic $p$-values are usually not heavily occupied and may carry information. For example, especially for smaller asteroids, the small $p$-values may also indicate irregular shapes (and the shadowing effects of nonzero solar phase angles) or an otherwise increased "noise level" due to systematic and modelling errors. In any case, the CDF method is meant to give a quick overview of a population instead of a detailed portrait, so trying to extract information via prior constraints is not a key concept here. An abundance of small $p$-values may also indicate that the data are simply not sufficient or otherwise suitable for a reliable result, so such a warning should not be suppressed by regularization.

\section{Simulations}

To assess the performance of our method with actual data and to check the effect of the simplified model, we perform several simulations. In Sect.\ 4.1, we explain the setup for our simulations and experiment with the synthetic data, giving graphical presentations of both actual and computed $(p, \beta)$ solutions. In Sect.\ 4.2, we discuss ways to apply a post-solution ``deconvolution'' in order to correct the systematic errors in the solution; the corrected solutions are graphically presented as well.

\subsection{Synthetic data}

Since the modelling errors and sampling effects in CDF construction dominate over the noise of original brightness data and the CDF errors do not show as signal noise, no standard error estimates are available. Therefore, the only way to test the reliability of our method is via simulations. In our setup, we utilize synthetic data for brightness measurements and attempt to reconstruct the $(p, \beta)$ distributions. In the simulations with the synthetic data, we use the same geometries (i.e. the direction vectors $\vec{e}_{\sun}$ and $\vec{e}_{\earth}$ as seen from the asteroid-fixed frame) and measurement time information that are used in the real asteroid databases, such as the Lowell Observatory database \citep{Bowell_et_al_2014}, the Asteroid Lightcurve Data Exchange Format (ALCDEF; \citet{Warner_et_al_2011_Save_the_lightcurves}), the Uppsala Asteroid Photometric Catalogue (UAPC; \citet{uapc01}, \citet{uapc02}) and the Wide-Field Infrared Survey Explorer (WISE; \citet{Mainzer_et_al_2011_Preliminary_results_from_NEOWISE}).

In our forward model, we take asteroid models from DAMIT  and apply basic transformations such as stretching on them to obtain a desired shape distribution with a large number of objects. For
DAMIT shapes, the concept of elongation is no longer as well defined as for ellipsoids,
but we estimate $p = b/a$ simply by choosing $a$ to be the longest diameter in the equatorial
$xy$-plane, and $b$ the width in the corresponding orthogonal direction. In this way, we generate a $(p, \beta)$ distribution with one peak by choosing suitable values of $p$ and $\beta$ for the objects.

\begin{figure}[!ht] \centering \includegraphics[width=0.4\textwidth]{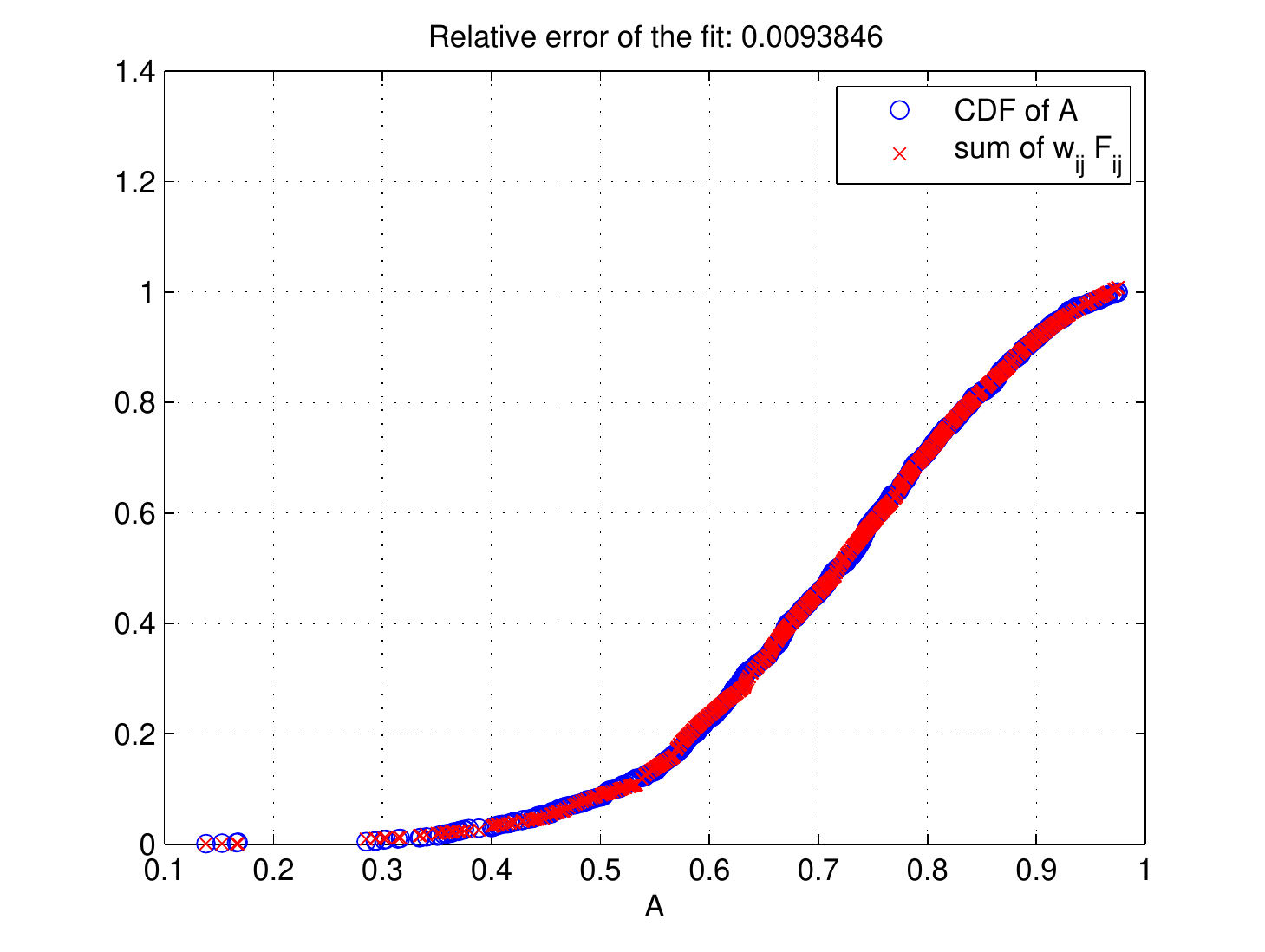} \caption{Function series $\sum_{ij} w_{ij} F_{ij}$ from Eq.\ \eqref{eq:lineq01} plotted in the same figure with the CDF $C(A)$ from data. The minimal error in the fitting should be noted.} \label{fig:Mock-fit} \end{figure}

For computing the brightnesses of the synthetic asteroids, we use a combination of the Lommel--Seeliger and Lambert scattering laws as in \citet{genproj}. We also add random perturbations to $L$ to simulate noise. When the brightness function has been computed, $\eta$ can be obtained using the discrete approximation of $\Delta (L^2)/\langle L^2\rangle$ from the available synthetic data points close enough in time to depict one rotation in a fixed geometry. Then, we get $A$ from Eq. \eqref{eq:eta-A}, and thus, the CDF of $A$: that is, $C(A)$. Other scattering models such as Hapke's could be used as well, but this represents only small brightness changes to separate objects and is thus not relevant to the collective results here. In fact, asteroid models in DAMIT are mostly constructed using the combined Lommel--Seeliger and Lambert law, so this choice reproduces the typical observed asteroid brightnesses best in this simulation. 

In the inverse problem, we attempt to reconstruct the original distribution. First of all, the competence of our method depends on how close our obtained distribution is to the original one. We can check numerically how well the function series $\sum_{ij} w_{ij} F_{ij}(A)$ of Eq.\ \eqref{eq:lineq01} converges to the CDF of $A$, $C(A)$, by computing the relative error:
\begin{equation}
\frac{\norm{C(A) - \sum_{ij} w_{ij} F_{ij}(A)}}{\norm{C(A)}} \stackrel{\text{Eq.\ } \eqref{eq:lineq02}}{=} \frac{\norm{C-Mw}}{\norm{C}} .
\label{eq:relerr}
\end{equation}
Figure\ \ref{fig:Mock-fit} depicts a typical fit of the analytical basis functions $F_{ij}$ to the data created with complex shapes and sampled more sparsely than implicitly assumed by the CDF integrals. We can see that the model usually fits CDF data perfectly, so the analytical basis functions provide a very good set despite the crudeness of the model approximations. The main question is thus the accuracy of the result rather than the explainability of the model.

In our simulations with synthetic data, we generate populations with a single $(p, \beta)$ peak in their joint distribution, and we attempt to reconstruct this peak. Each asteroid can have multiple brightness measurements, and we require at least five observation points for a valid estimate of the variation observable $\eta$. From the results of the simulations, we have found this to be the typical minimum number of data points for sampling one rotation of the target. This is also simple to estimate analytically by considering the possible permutations of random samples of a boxcar-shaped sinusoidal signal: then the average error of  $\eta$ drops fast from the $50\%$ of two sample points to close to $10 \%$ with five or six points. Such error levels already fit well in the total error budget. Numerical examples of lightcurves give similar results.

In order to obtain enough observations of $\eta$ and thus accurate distributions, we use populations of 1000 asteroids. This is a realistic population size, as real major databases contain $\propto 10^3$--$10^4$ objects. Smaller populations start to suffer from too sparse sampling of geometries. Also, the systematic errors caused by the inaccurate assumptions of the model and the sparsity of rotational phases in $\eta$ estimation are best counteracted by the averaging effect of a large number of samples. For the bins, $p_i \in [0, 1]$ and $\beta_j \in [0, \pi/2]$, where we have selected $i=1$, $\ldots$, $20$ and$j=1$, $\ldots$, $29$, with a random point near the centre of each equally spaced bin to represent its $\beta$ or $p$ value. This way, every bin is about $0.05 \times 0.05$ units in size, and the computation of the inverse problem is fast enough.

\begin{figure}[!ht] \centering \includegraphics[width=0.242\textwidth]{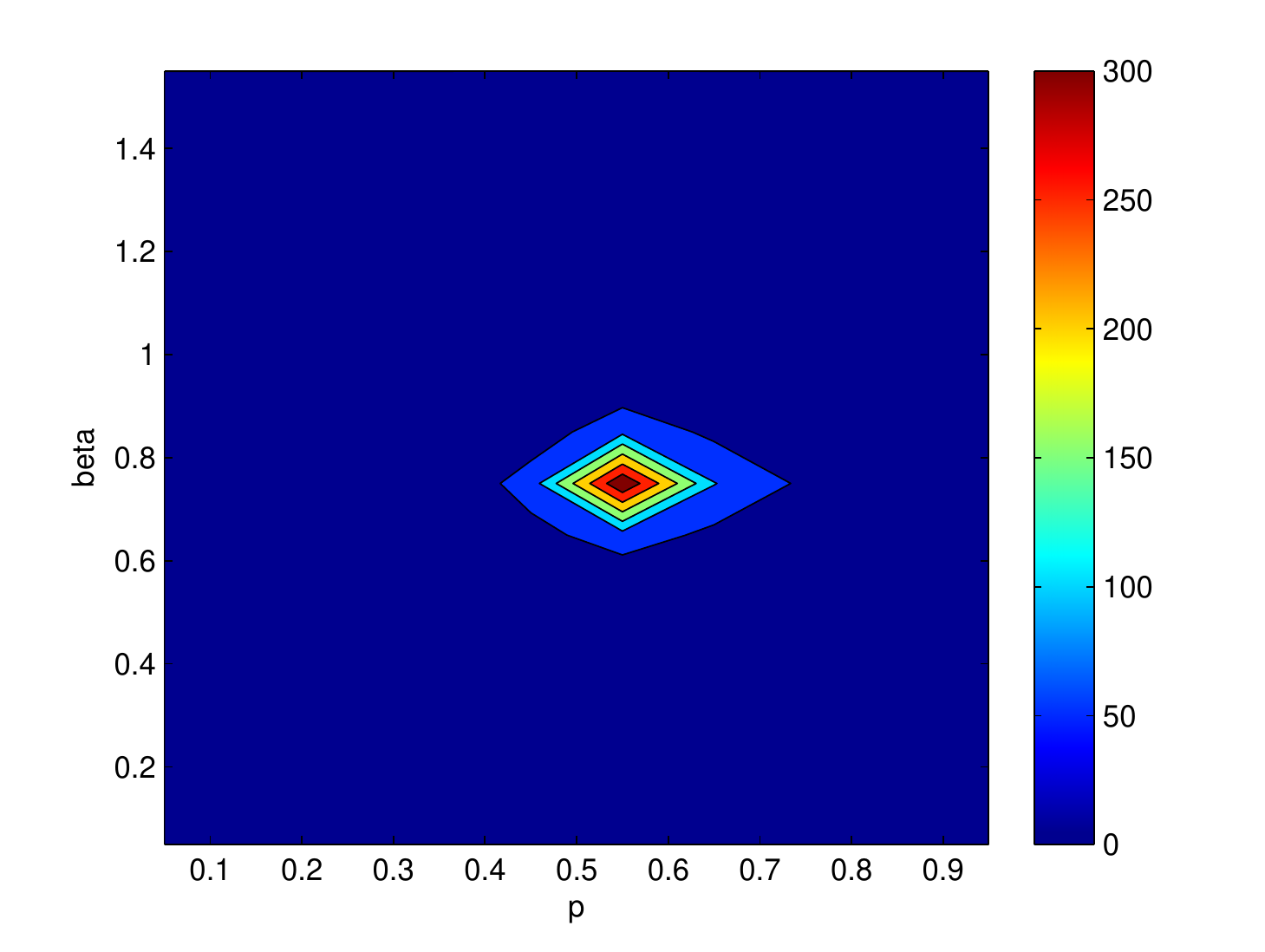} \includegraphics[width=0.242\textwidth]{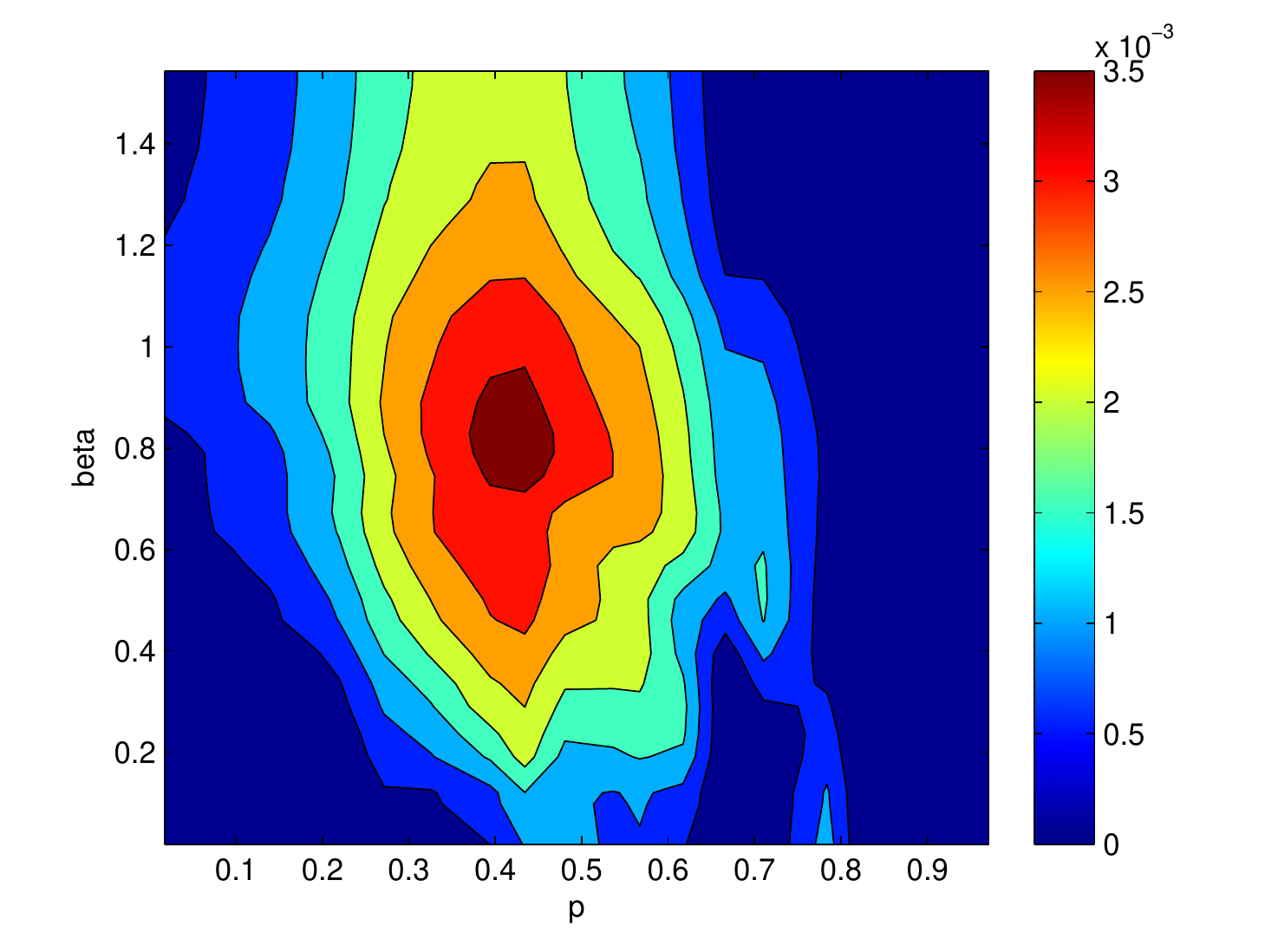} \includegraphics[width=0.242\textwidth]{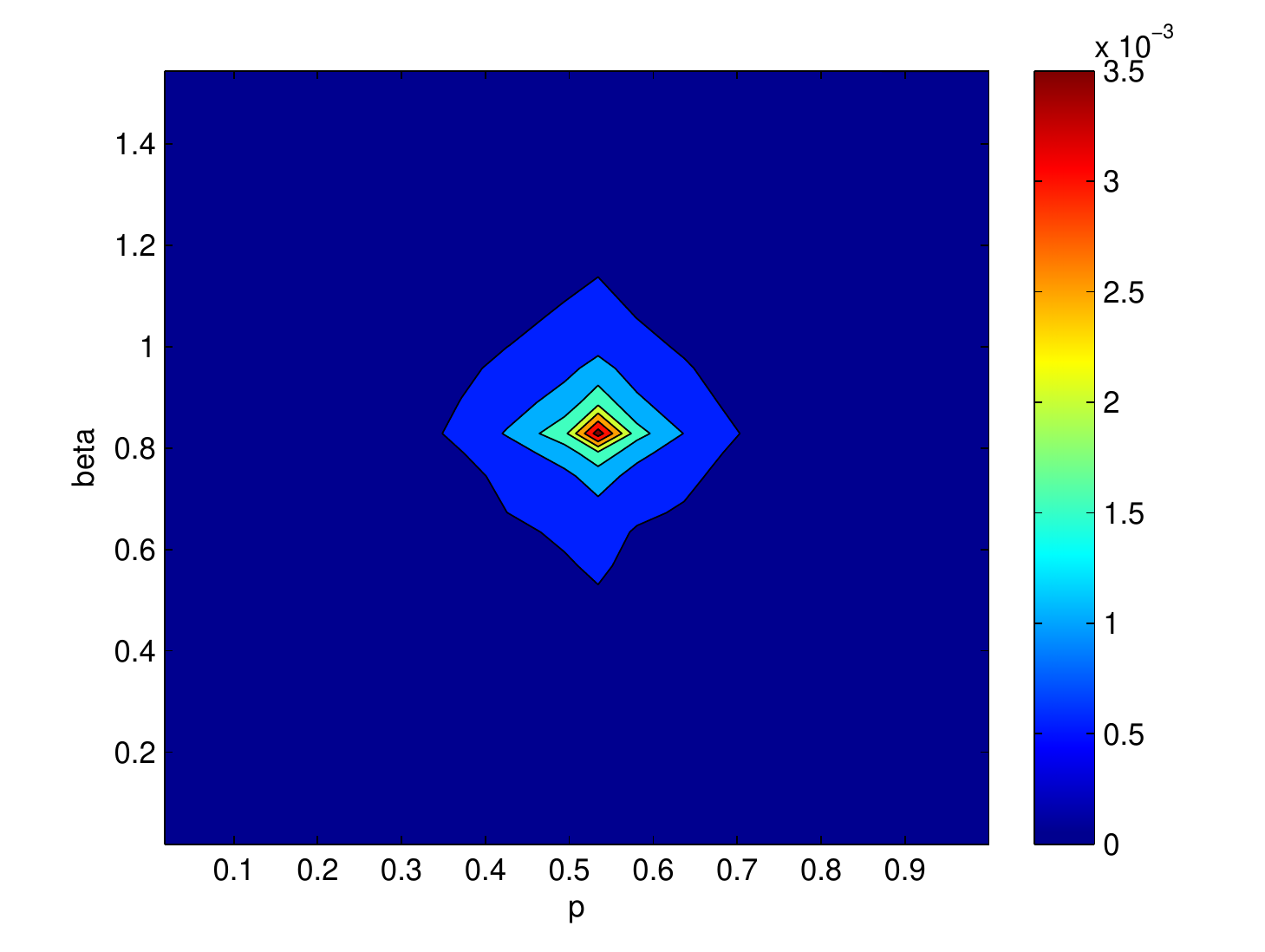} \caption{Actual joint distribution of $(p, \beta)$ (top left) of synthetic asteroids compared to the solution of the WISE-based inverse problem (top right). The colours depict the occupation number of each $(p, \beta)$-cell (on arbitrary scales). The absolute value of the ecliptic latitude of the spin axis decreases from bottom (perpendicular to the ecliptic plane) to top (in the ecliptic plane), and the shape elongation decreases from left (thin cigar) to right (sphere). On the bottom is the solution of the inverse problem after applying deconvolution for correction.} \label{fig:synth-medmed} \end{figure}

\begin{figure}[!ht] \centering \includegraphics[width=0.242\textwidth]{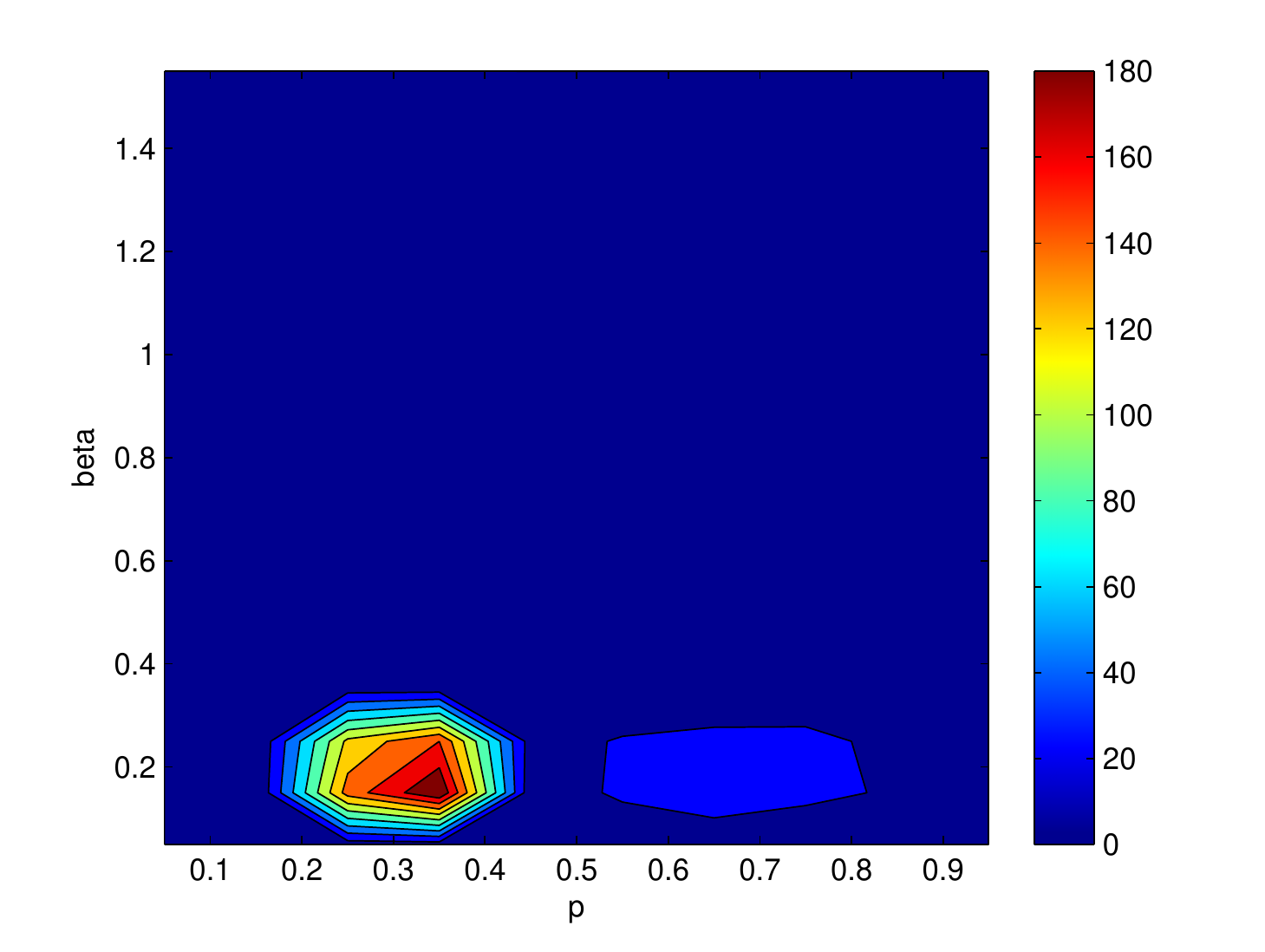} \includegraphics[width=0.242\textwidth]{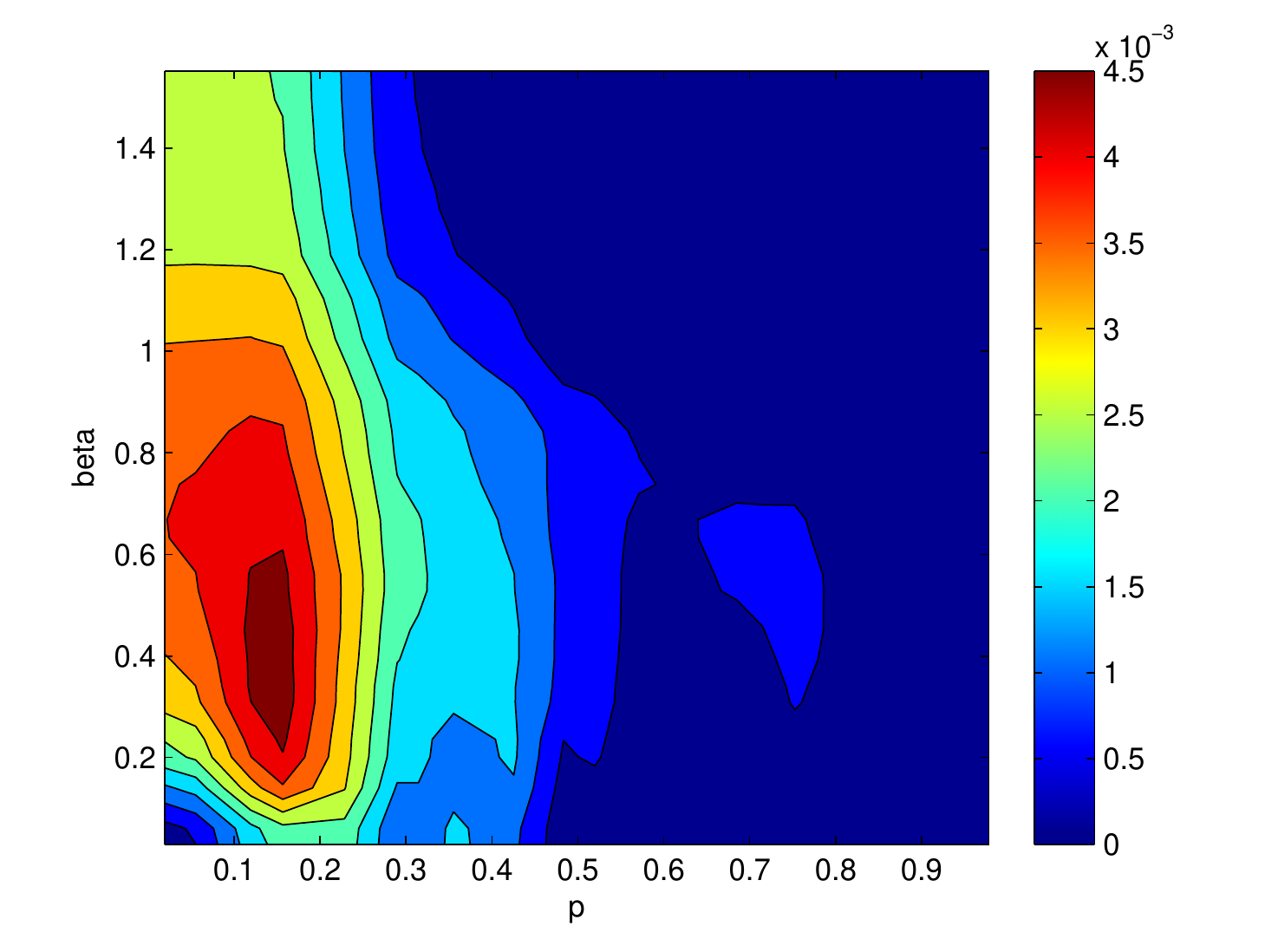} \includegraphics[width=0.242\textwidth]{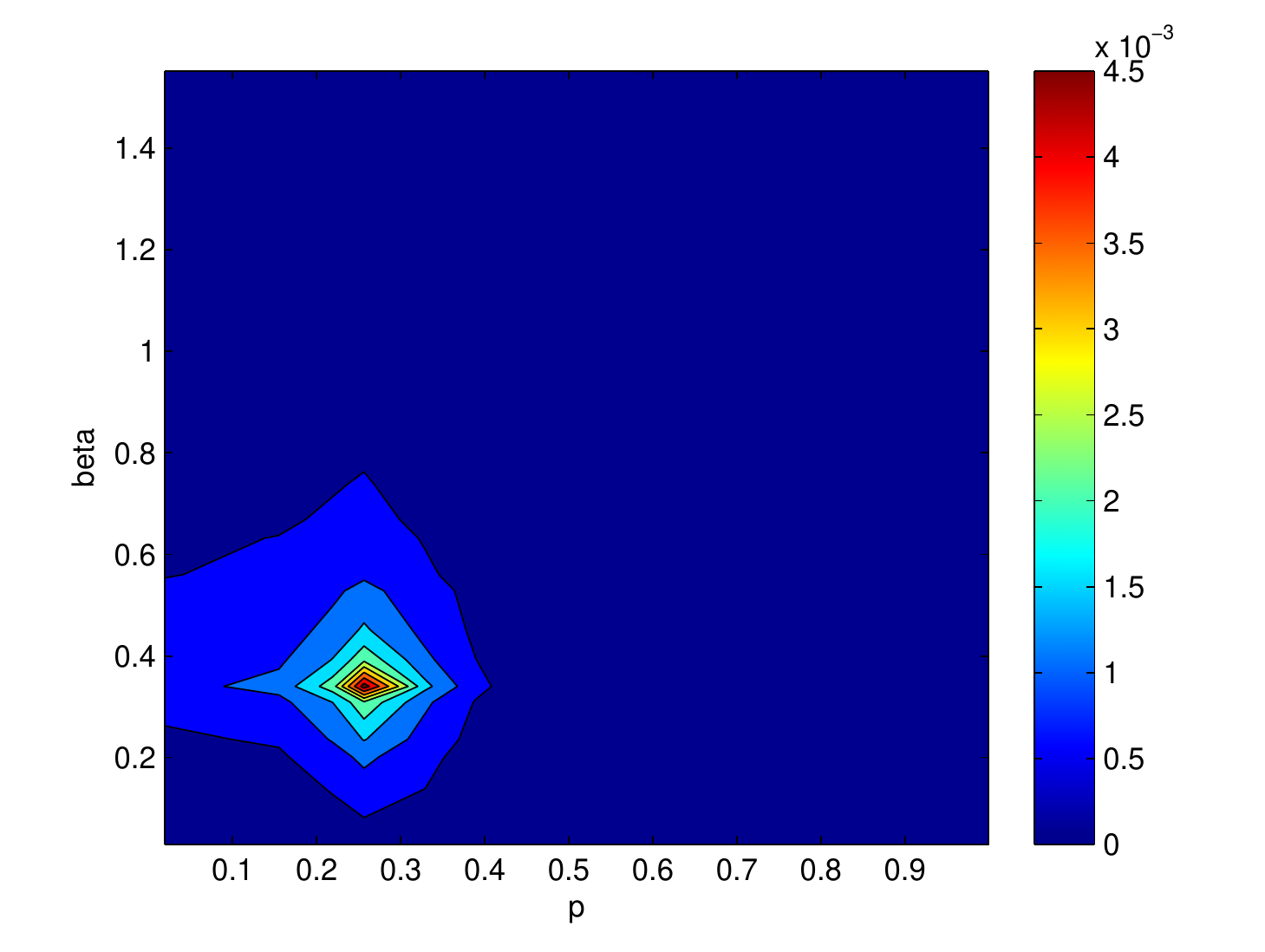} \caption{Actual joint distribution of $(p, \beta)$ (top left) of synthetic asteroids compared to the solution of the WISE-based inverse problem (top right). Here we have tested how accurately the solution is obtained if the peak of both $p$ and $\beta$ distributions is low. On the bottom is the solution with deconvolution added.} \label{fig:synth-lowlow} \end{figure}

\begin{figure}[!ht] \centering \includegraphics[width=0.242\textwidth]{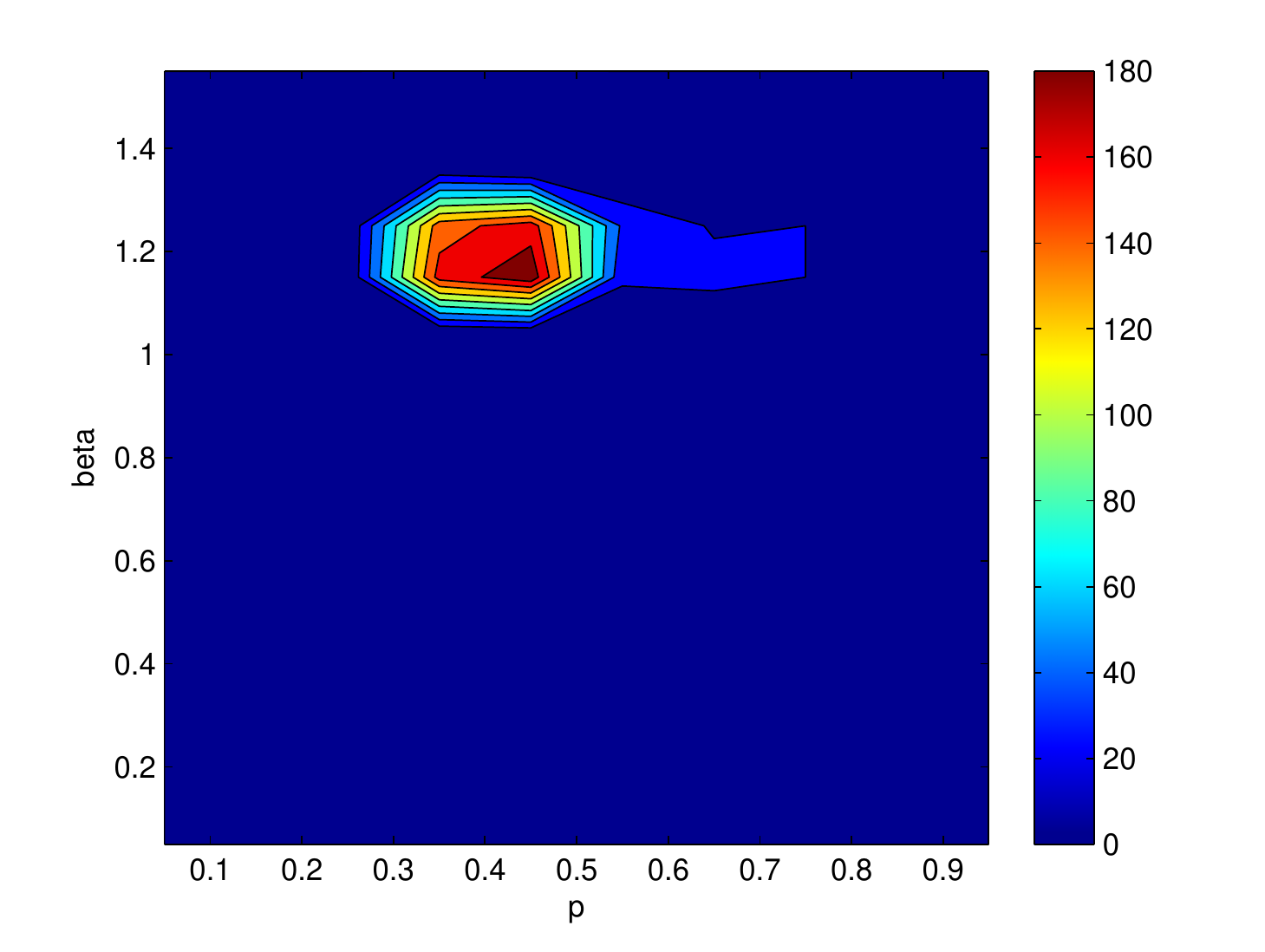} \includegraphics[width=0.242\textwidth]{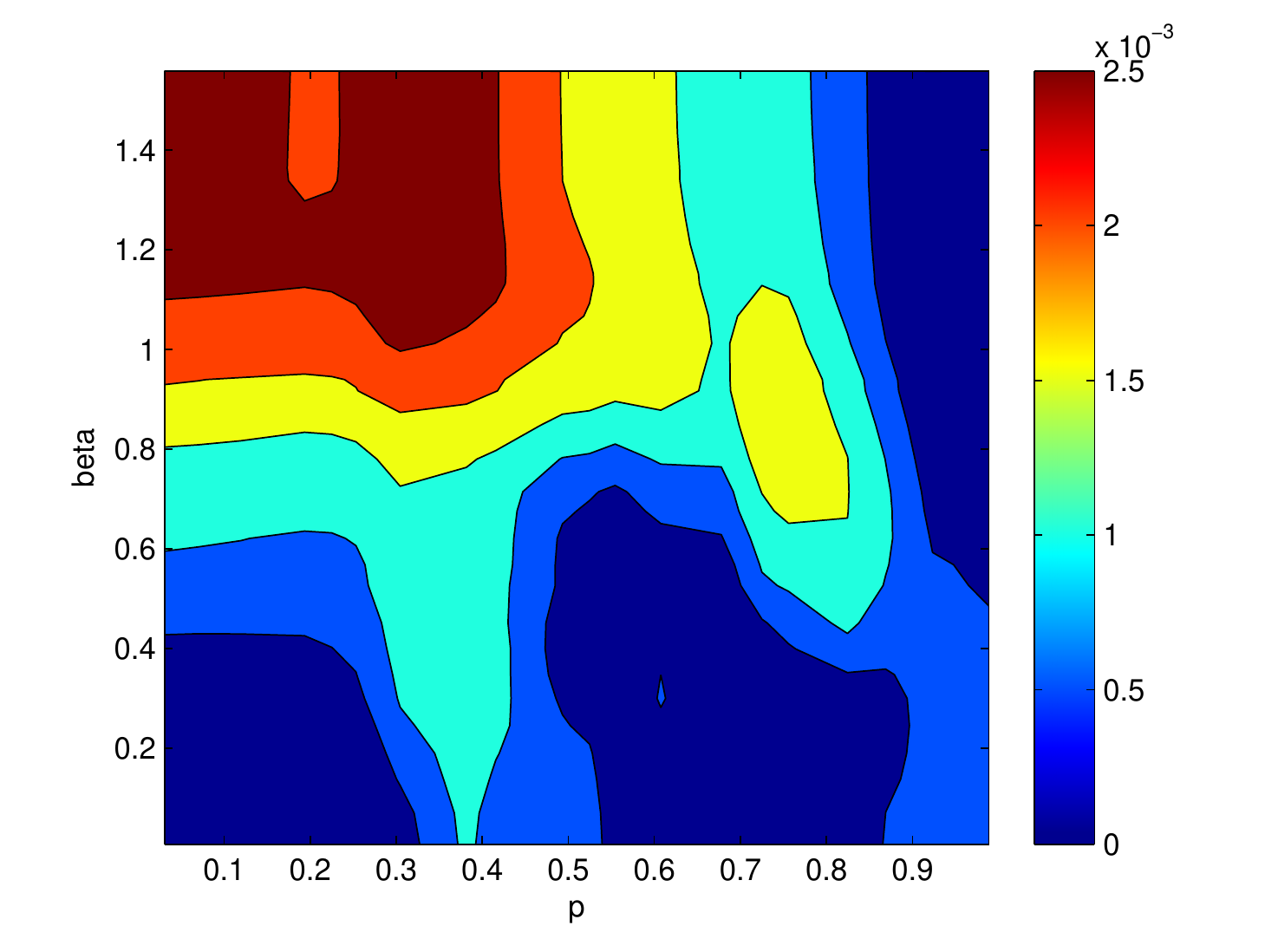} \includegraphics[width=0.242\textwidth]{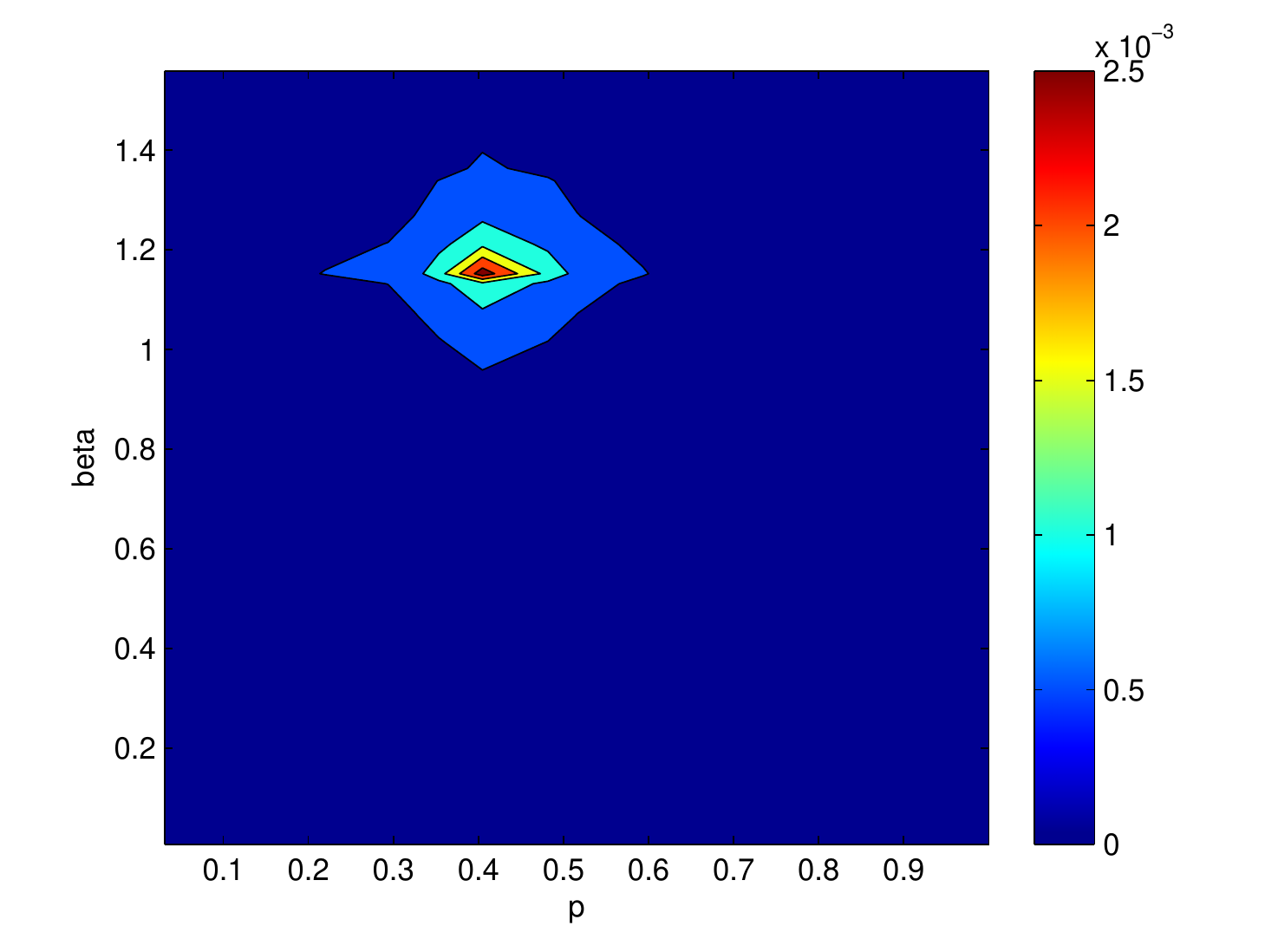} \includegraphics[width=0.242\textwidth]{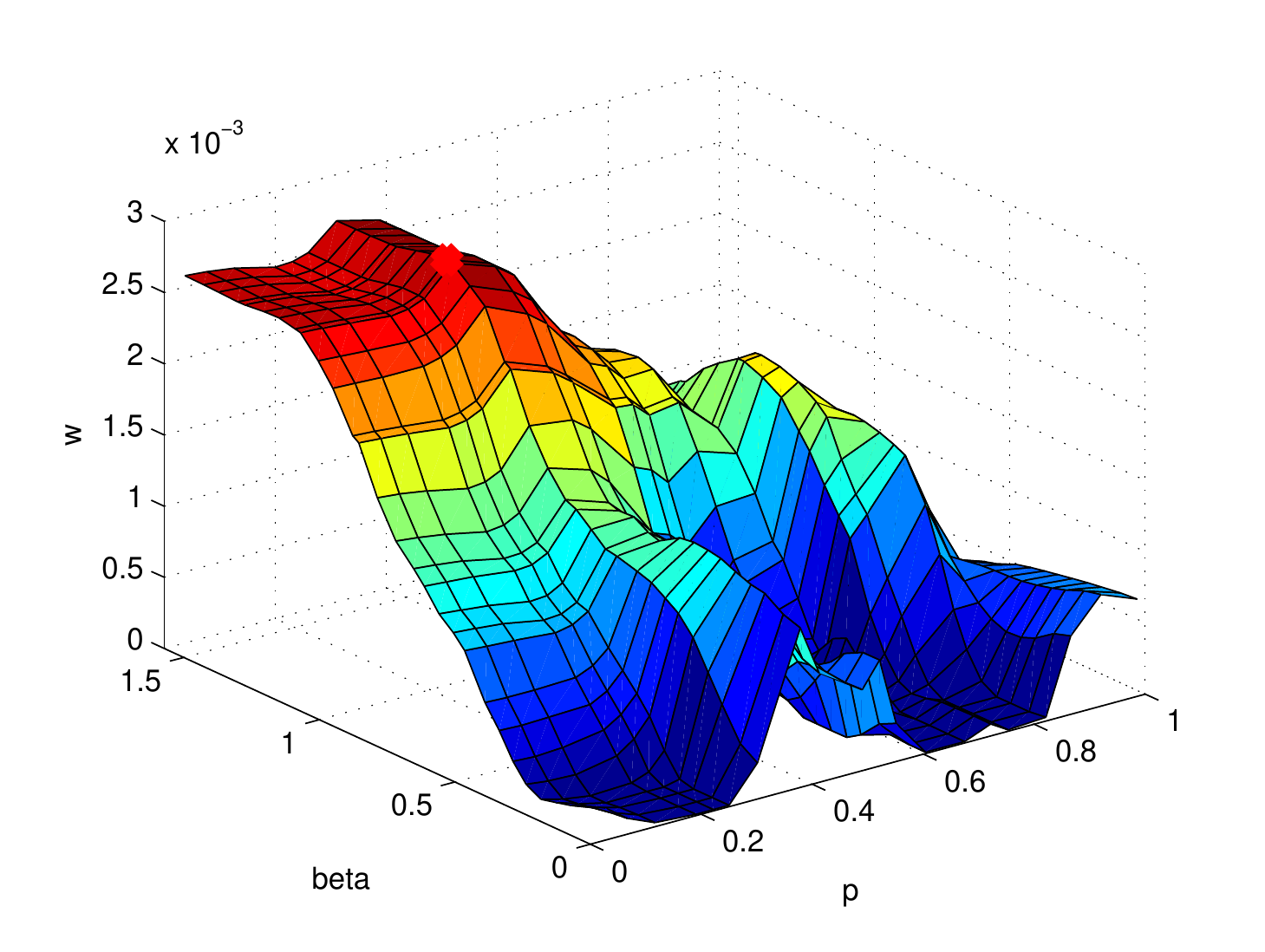} \caption{Actual joint distribution of $(p, \beta)$ (top left) of synthetic asteroids compared to the solution of the WISE-based inverse problem (top right). To make the solution plot more easily readable, we have plotted it from another perspective on the bottom right, with the $z$-axis depicting the weights $w$ of each $(p, \beta)$ bin. Here we have tested how accurately the solution is obtained if the peak of the distributions is low for $p$ and high for $\beta$. On the bottom left is the solution with deconvolution.} \label{fig:synth-lowhigh} \end{figure}

\begin{figure}[!ht] \centering \includegraphics[width=0.242\textwidth]{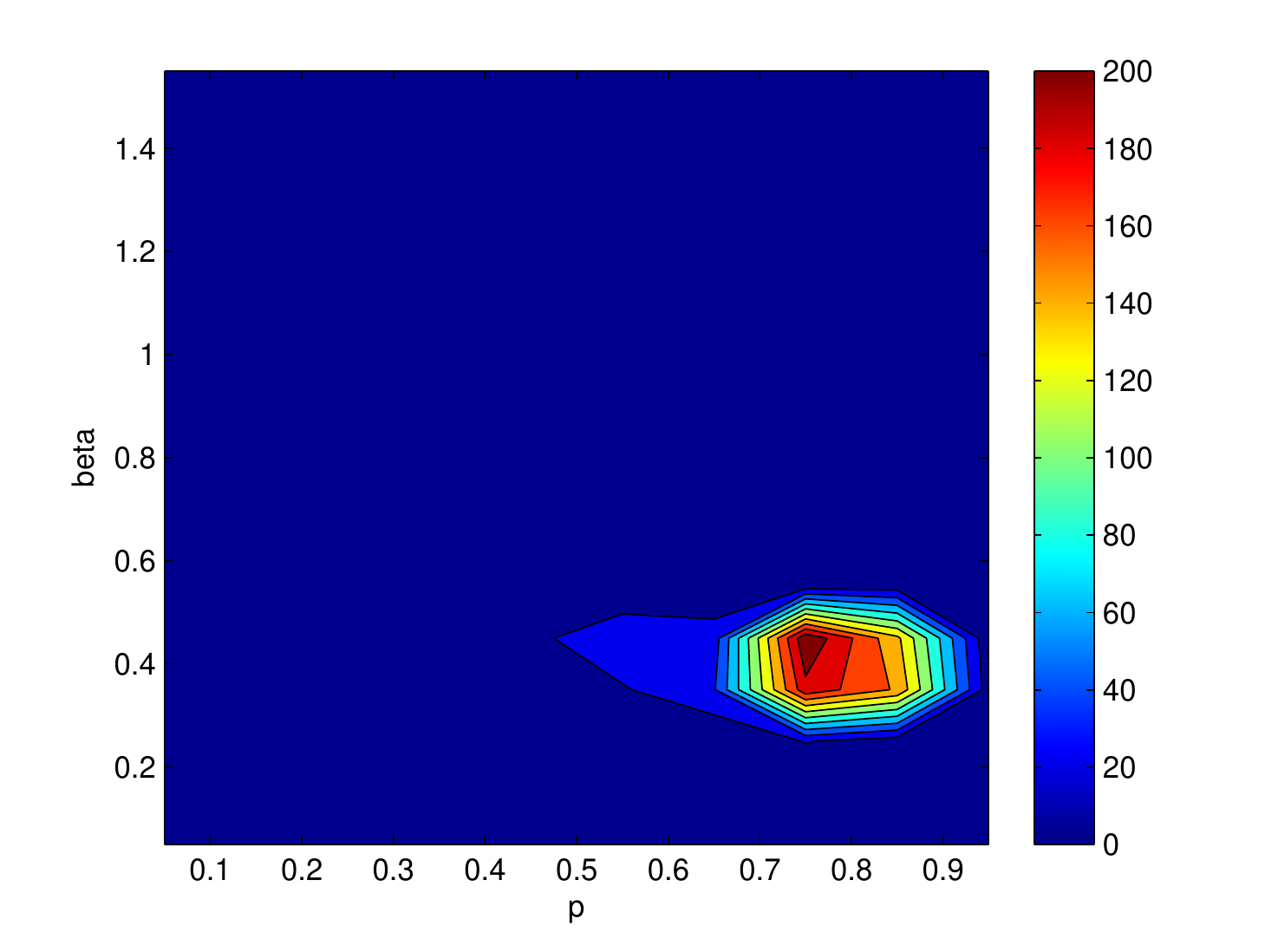} \includegraphics[width=0.242\textwidth]{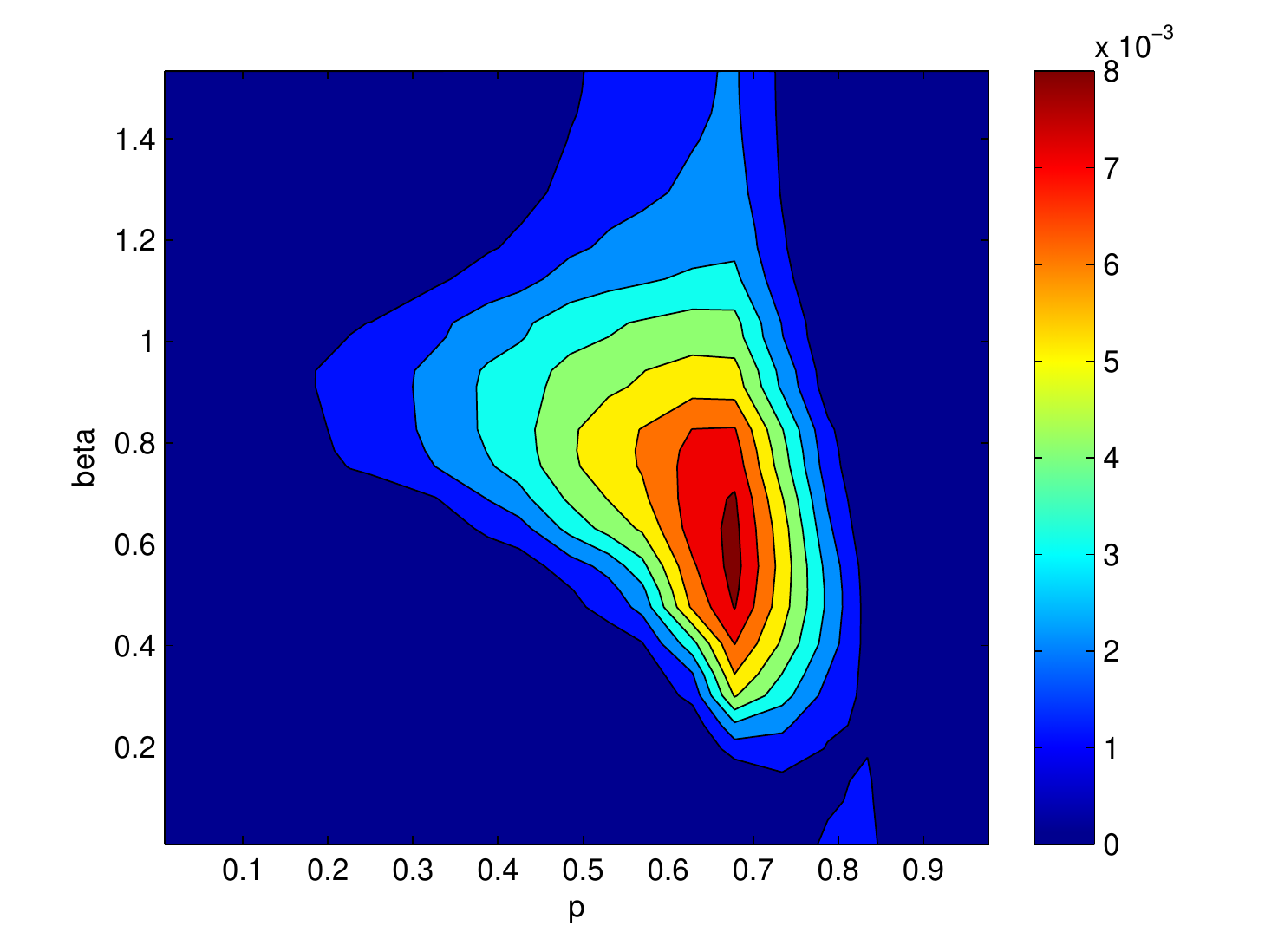} \includegraphics[width=0.242\textwidth]{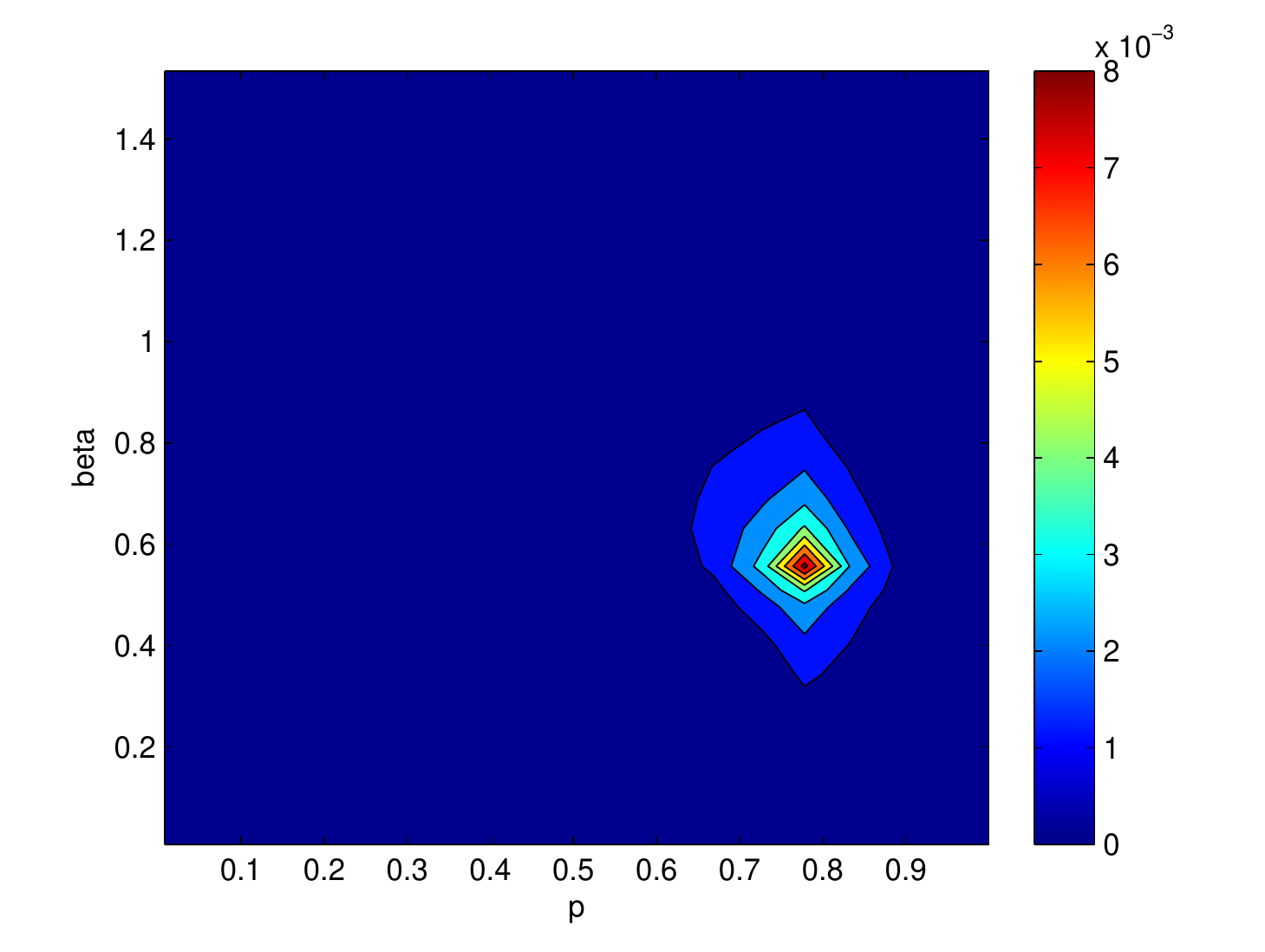} \caption{Actual joint distribution of $(p, \beta)$ (top left) of synthetic asteroids compared to the solution of the WISE-based inverse problem (top right). Here we have tested how accurately the solution is obtained if the peak of the distributions is high for $p$ and low for $\beta$. On the bottom is the solution with deconvolution.} \label{fig:synth-highlow} \end{figure}

\begin{figure}[!ht] \centering \includegraphics[width=0.242\textwidth]{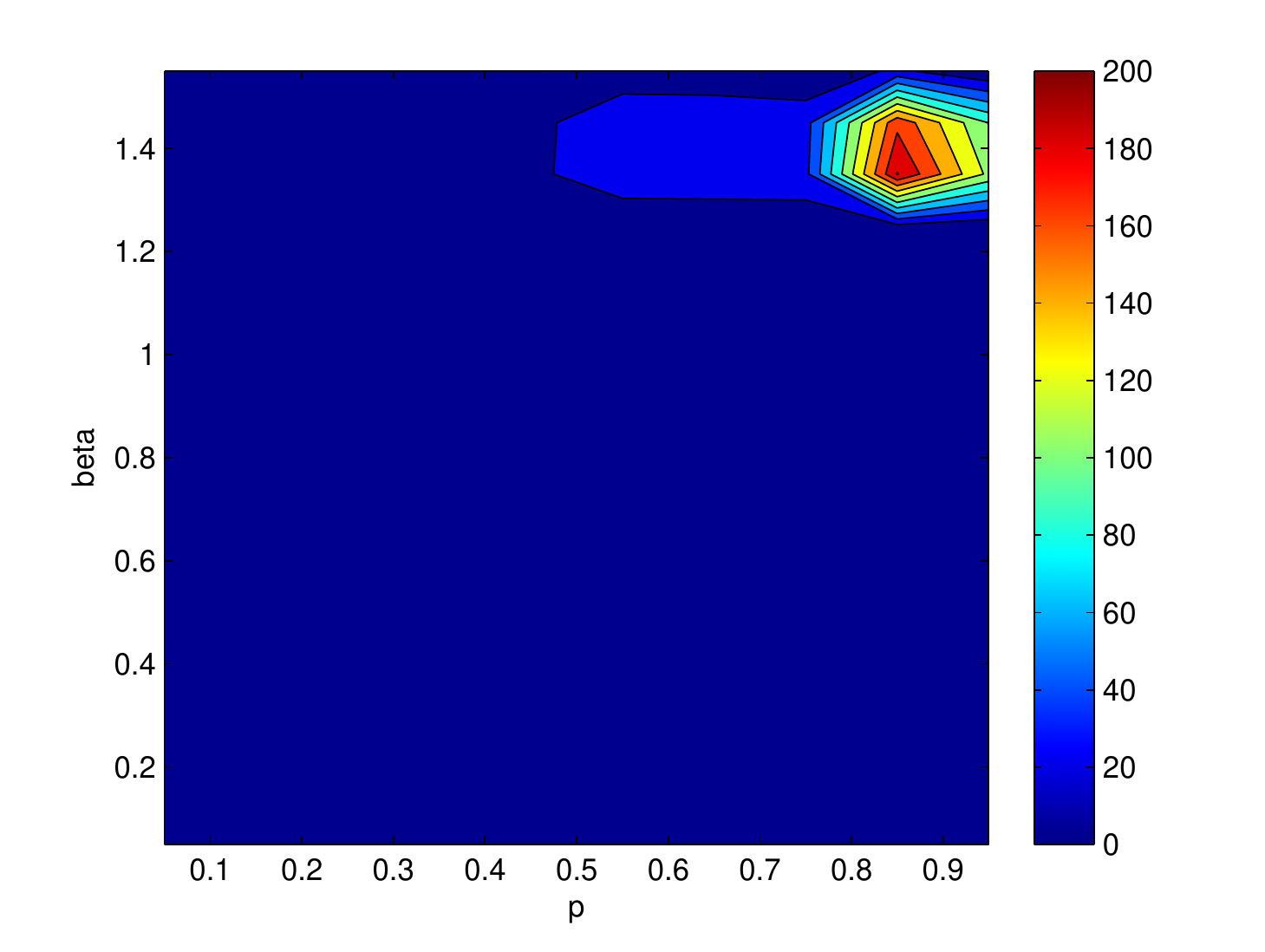} \includegraphics[width=0.242\textwidth]{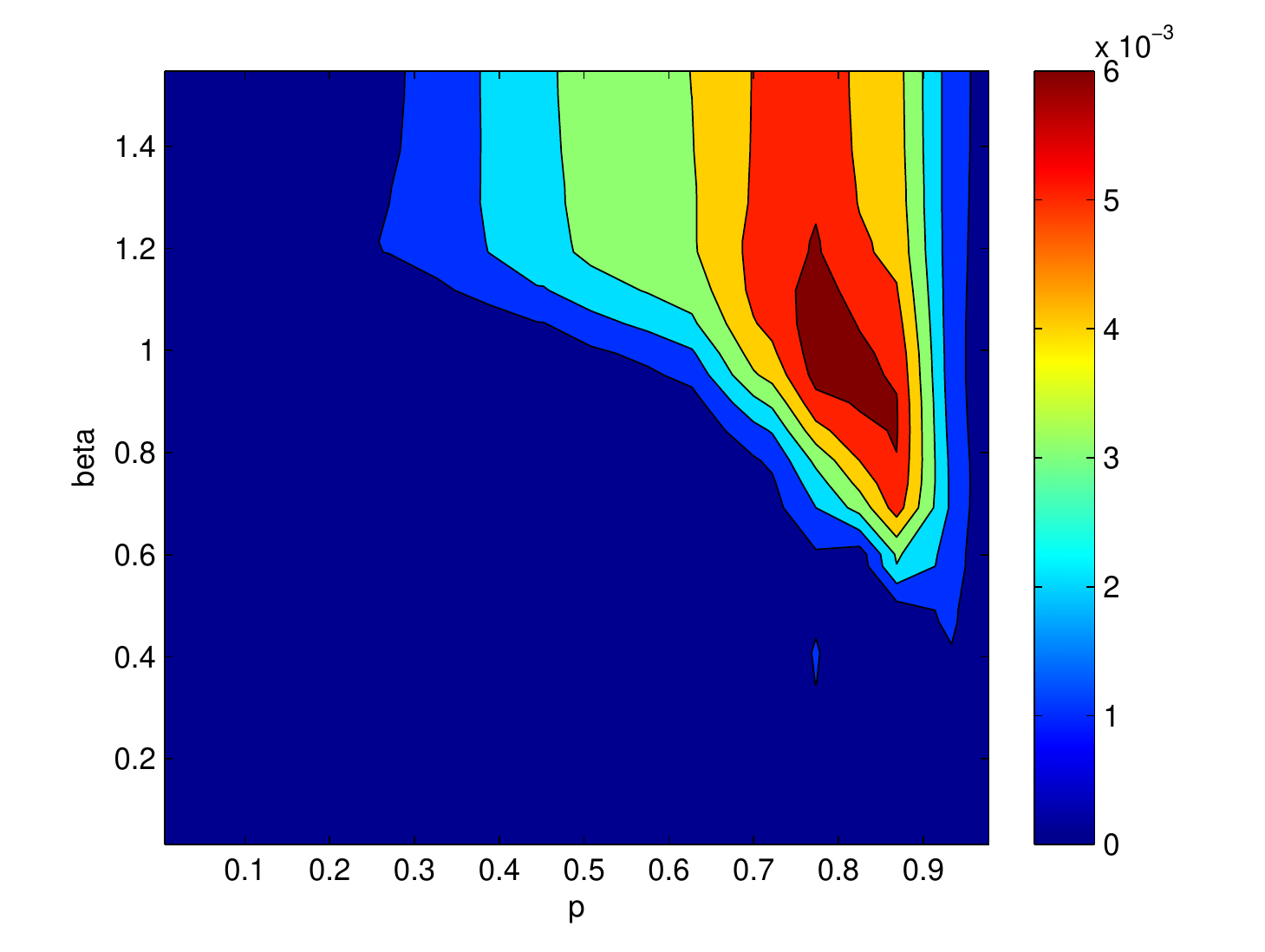} \includegraphics[width=0.242\textwidth]{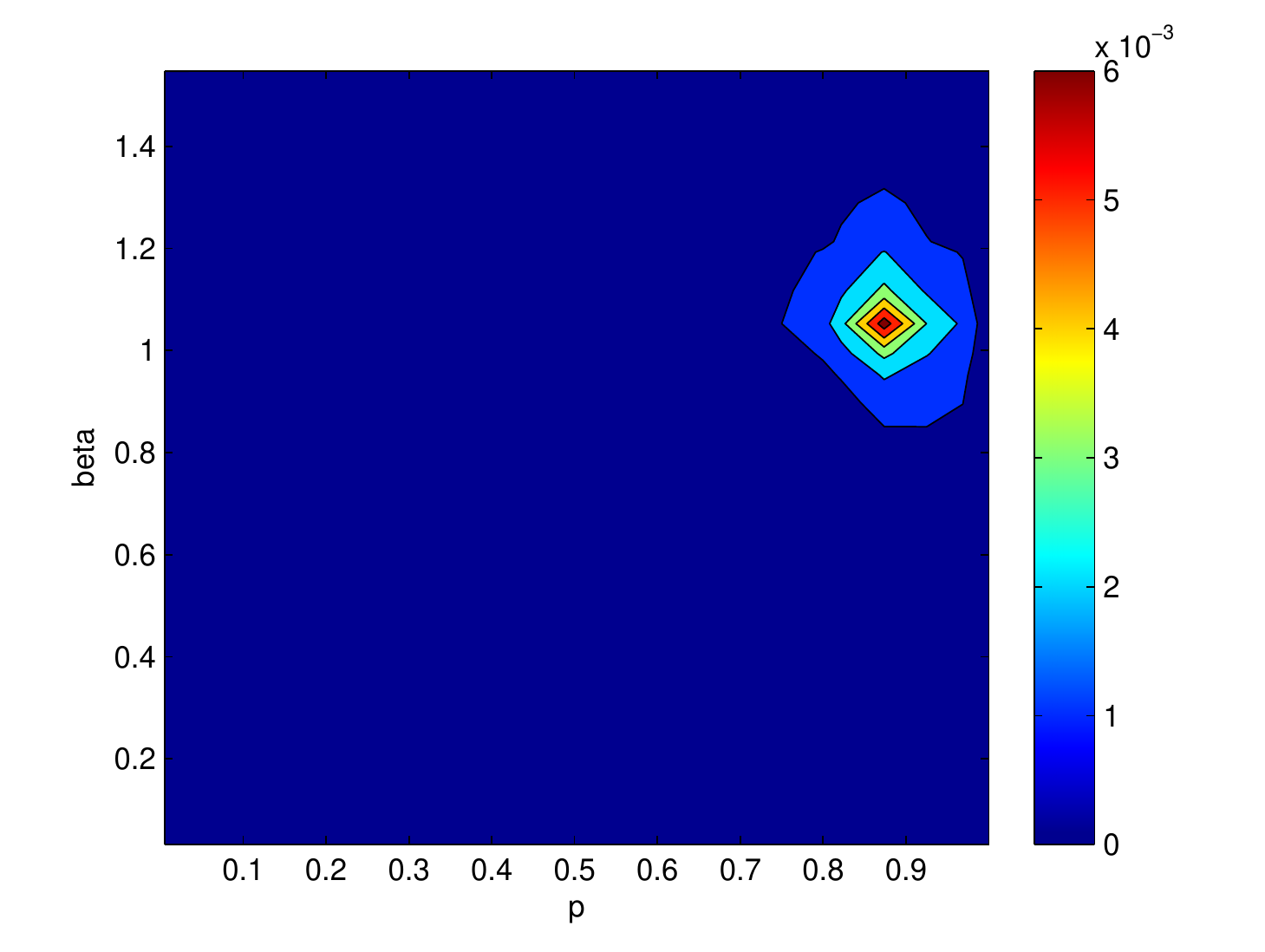} \caption{Actual joint distribution of $(p, \beta)$ (top left) of synthetic asteroids compared to the solution of the WISE-based inverse problem (top right). Here we have tested how accurately the solution is obtained if the peak of both $p$ and $\beta$ distributions is high. On the bottom is the solution with deconvolution.} \label{fig:synth-highhigh} \end{figure}

We have plotted some all-round cases of the $(p, \beta)$ distribution of the forward model and the solution distribution of the inverse problem (from data mimicking WISE) in Figs.\ \ref{fig:synth-medmed}--\ref{fig:synth-highhigh}. In the forward model, the peak of the $(p, \beta)$ distribution has been placed in the middle, bottom left, top left, bottom right, and top right positions in the $(p, \beta)$ plane, respectively. In Fig.\ \ref{fig:synth-medmed}, we notice that the approximate location of the $(p, \beta)$-peak is correct, but the solution spreads when moving away from the peak, particularly when moving towards $\pi/2$ (spin direction in the ecliptic plane). In addition, the peak is too much to the left in $p$-axis (towards more elongated bodies). The same phenomenon can be observed in the other figures as well. In Fig.\ \ref{fig:synth-lowhigh}, when the peak was located in the top left corner of the $(p, \beta)$ plane (elongated bodies in the ecliptic plane), the contour looked visually messy every time. Hence, we included an additional plot of the solution of the inverse problem in the $(p, \beta, \mathrm{DF}(p, \beta))$ coordinates. From the three-dimensional perspective, we notice that the peak of the shape elongation is once again too far in the left in $p$-axis, and the solution spreads when moving away from the peak. Indeed, these errors are systematic, and they occur in a solution every time. The $p$-shift is inevitable: even in the absence of noise, near-spherical targets with $p=1$ do not portray a completely flat lightcurve because of local shape irregularities. In addition, Figs. \ref{fig:synth-lowlow}, \ref{fig:synth-highlow}, and \ref{fig:synth-highhigh} show a trend of the peak of the $\beta$ solution to move slightly towards the middle (away from the $\beta=0$ and $\beta=\pi/2$ ends). The error is common but does not occur every time. In general, the errors are encountered because of both modelling errors and noisy measurements. They are rather regular and predictable, and therefore, it is possible to formulate a post-solution correction in order to revise the solution distribution.

\subsection{Correction in $(p,\beta)$-plane}

The "deconvolution" of the noisy solution "image" in the $(p,\beta)$-plane is a visual aid based on experiments performed on the synthetic data. With the help of simulations, we were able to acquire a good understanding of how much our computational solution typically differs from the actual distribution when one assumes that there is a dominant peak in the latter. This way, we could deduce the typical point-spread function of the solution in the plane. We introduce damping on bins further away from the peak of the centre of the solution. Then we move the values of $p$ a constant (fixed) step to the right: $p_i \to p_i + \Delta \mathcal{P}$. According to our simulations with the WISE database, the solution is typically shifted about 0.1 $p$-units to the left due to noise. Therefore, we choose $\Delta \mathcal{P} = 0.1$ when we use WISE data. The systematic error in $\beta$ direction is irregular, and there is no way to know whether the obtained $\beta$ is too small or too large. For $\beta$, we observed that the obtained distribution is usually accurate if the actual $\beta$ peak is somewhere near $\pi/4$ (usually when $\beta \in [0.5, 1]$), but if the actual peak is near the extreme end values $0$ or $\pi/2$, then the solution tends to shift the peak away from the extremes, towards the middle values.

We show the deconvoluted solution in the bottom figures of Figs.\ \ref{fig:synth-medmed}--\ref{fig:synth-highhigh} (bottom left picture in Fig.\ \ref{fig:synth-lowhigh}). The deconvolution has been used to correct the solution presented in the top right picture of the same figure. The corrected solution is close to the distribution shape of the forward model in the top left picture. 
We will apply deconvolution solely on the joint $(p, \beta)$ distribution. In order to reduce errors and loss of information, the marginal $p$ and $\beta$ distributions are presented without corrections. In their cases, the main point is that the peak of the $p$ distribution is usually slightly more to the right in the $p$-axis than in the obtained solution.

\section{Results from astronomical databases}

% Population sizes in WISE database
%
% Vesta: 1338
% Flora: 2849
% Nysa Polana: 2305
% Massalia: 154
% Gefion: 673
% Eunomia: 1972
% Maria: 945
% Koronis: 642
% Themis: 1709
% Hygiea: 1549
% Eos: 3323
% Alauda: 771
% Phocaea 1022

In this section, we plot distributions of different asteroid families and introduce a method for comparing such distributions.
The setup we use is very similar to the one used for synthetic data. We receive our data (geometries, brightness values, measurement times) from the WISE database. We downloaded the data from the Infrared Science Archive: Infrared Processing and Analysis Center (IRSA/IPAC archive\footnote{\url{http://irsa.ipac.caltech.edu/Missions/wise.html}.}) and used the same selection criteria as \cite{Ali-Lagoa_et_al_2014}. The combined ALCDEF \& UAPC lightcurve database (hereafter called simply ALCDEF) is also useful for various analyses, but its denser lightcurves (yielding improved $\eta$ estimates) do not really compensate for the larger number of objects in the WISE data that is crucial to the robustness of the statistical CDF approach, as discussed earlier. Moreover, the lightcurves sample well only asteroids with short rotation periods, because observations from different nights usually cannot be combined together because of poor or completely missing calibration. The large number of brightness variation samples, rather than the accuracy of the observable, is the main reason why the method can tolerate the crude underlying shape model. Even though the WISE data are in mid-infrared wavelengths (we used measurements at 12 and 22 $\mu$m), the $\eta$ derived for them is essentially the same as from the projected area since the infrared regime mainly causes a lag in the lightcurves (compared to visual data) that does not affect the brightness variation \citep{wise}.

Another possible rich source of asteroid photometry is the Lowell Observatory photometric database \cite{Bowell_et_al_2014}, which was used by \cite{Oszkiewicz_et_al_2011} and \cite{cib}. However, the large errors of photometric points of $\sim$0.1--0.2 mag would bring another source of systematic error into our model.

We consider one $\eta$-estimate to consist of measurements done within a three-day time window to keep the observing geometry sufficiently constant. We only accept estimates based on at least five measured values. In principle, our model requires the phase angle between the Sun and Earth to be close to zero degrees. However, adding this restriction would greatly reduce the number of brightness measurements we could use, and would eventually lead to a considerably lower number of $\eta$ values. According to simulations performed on synthetic data, the error caused by a non-zero phase angle is so small compared to other error sources that its effect is negligible. Therefore, we set a very liberal requirement that
        \begin{equation}
                \arccos( \vec{e}_{\sun} \cdot \vec{e}_{\earth} ) \le 30^{\circ} .
        \label{eq:ang-limit}
        \end{equation}

\subsection{Discussion about bias}

Before we move on to plotting asteroid families, we discuss some possible sources for biases.
The number of possible $\eta$-estimates varies between asteroids; if an asteroid yields $n$ estimates of $\eta$, we can formally give each of these estimates the weight $1/n$. However, if there are many estimates associated with some asteroids, there could be a bias, as the solution could be favouring such targets. We checked if the solution was affected if we only took one estimate for each target. There was no noticeable change from the situation when all estimates were considered, so we can conclude that there is no significant bias from the weights of individual asteroids when using large databases for a large number of objects. This underlines the safety in large numbers, so the method can be used even if we do not know which observation is from which asteroid and use all $\eta$ estimates "blindly".

\begin{figure}[!ht] \centering \includegraphics[width=0.49\textwidth]{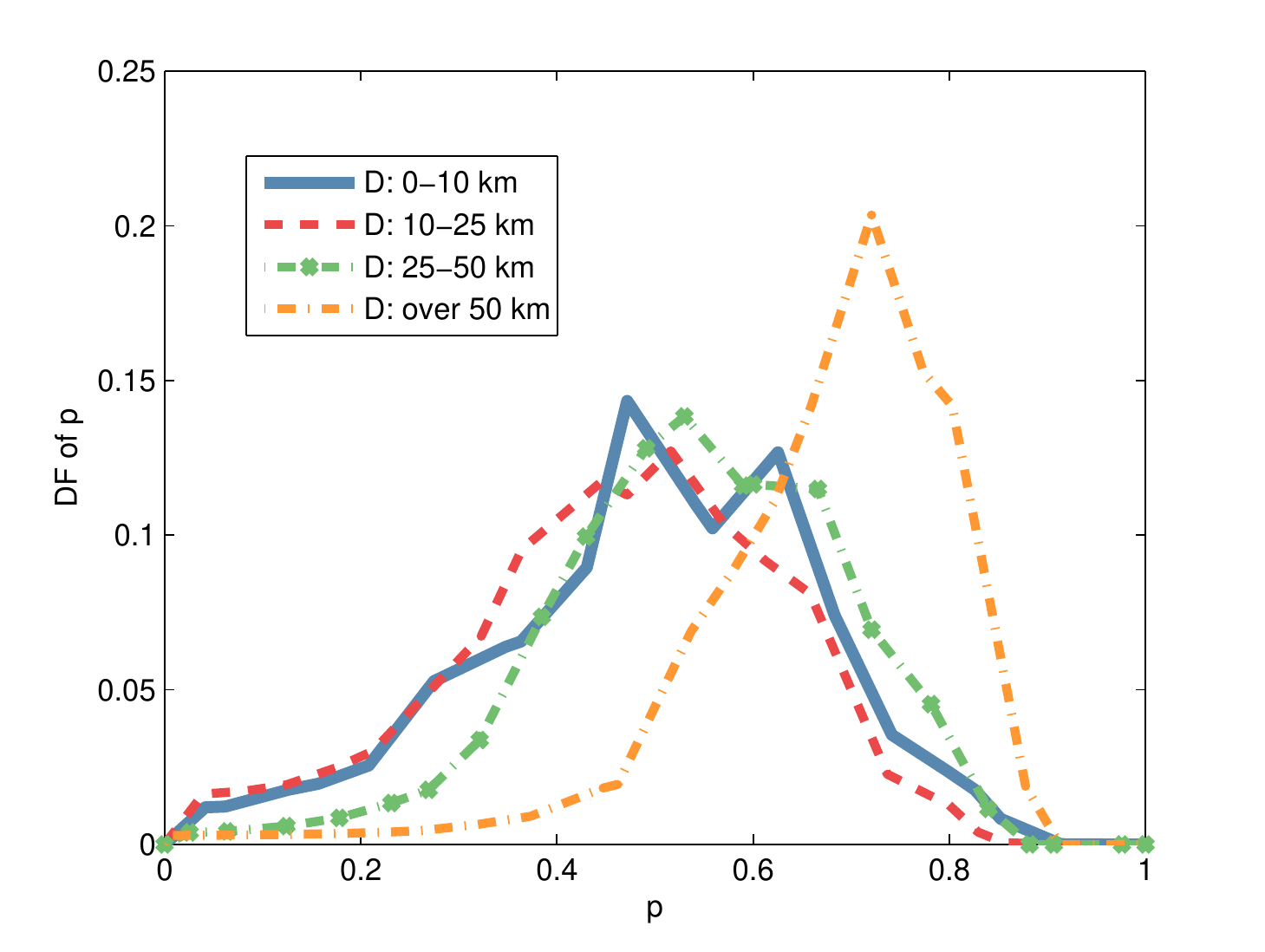} \caption{Comparison of the marginal DFs of the shape elongation $p$ for asteroids of different sizes from WISE data. The vertical axis depicts the occupation number of each $p$-slot (on an arbitrary scale). The shape elongation decreases from left (thin cigar) to right (sphere).} \label{fig:WISE-sizecomp01} \end{figure}

As we know from asteroid lightcurves, large asteroids are generally more spherical than small asteroids. This was also shown by \cite{cib}, for example. We checked this result by comparing WISE subpopulations of different sizes. We divided the WISE asteroids into four subpopulations: ones with diameter $D < 10 \ \mathrm{km}$ ($\approx$ 65,000 bodies), $10 \ \mathrm{km} \le D < 25 \ \mathrm{km}$ ($\approx$ 6000 bodies), $25 \ \mathrm{km} \le D < 50 \ \mathrm{km}$ ($\approx$ 1000 bodies), and $D \ge 50 \ \mathrm{km}$ ($\approx$ 1000 bodies). While the population sizes are noticeably different (the number of small asteroids clearly surpasses the number of large ones), the subpopulations were selected so that each of them would have a sufficiently large sample size in order to acquire reliable results. The obtained distributions confirmed that large asteroids indeed tend to be more spherical (see Fig.\ \ref{fig:WISE-sizecomp01} for comparison). Our result is considerably different from the one obtained by \citet{mcneill}, where the peak of the $p$ distribution for $D < 8 \ \mathrm{km}$ was located at a near-spherical value of $b/a=0.85$. Of course, the distribution tail of small $p$-values of our result for $D < 10 \ \mathrm{km}$ is more due to systematic and model errors (especially irregular shapes) than to actually very elongated bodies.

\begin{figure}[!ht] \centering \includegraphics[width=0.49\textwidth]{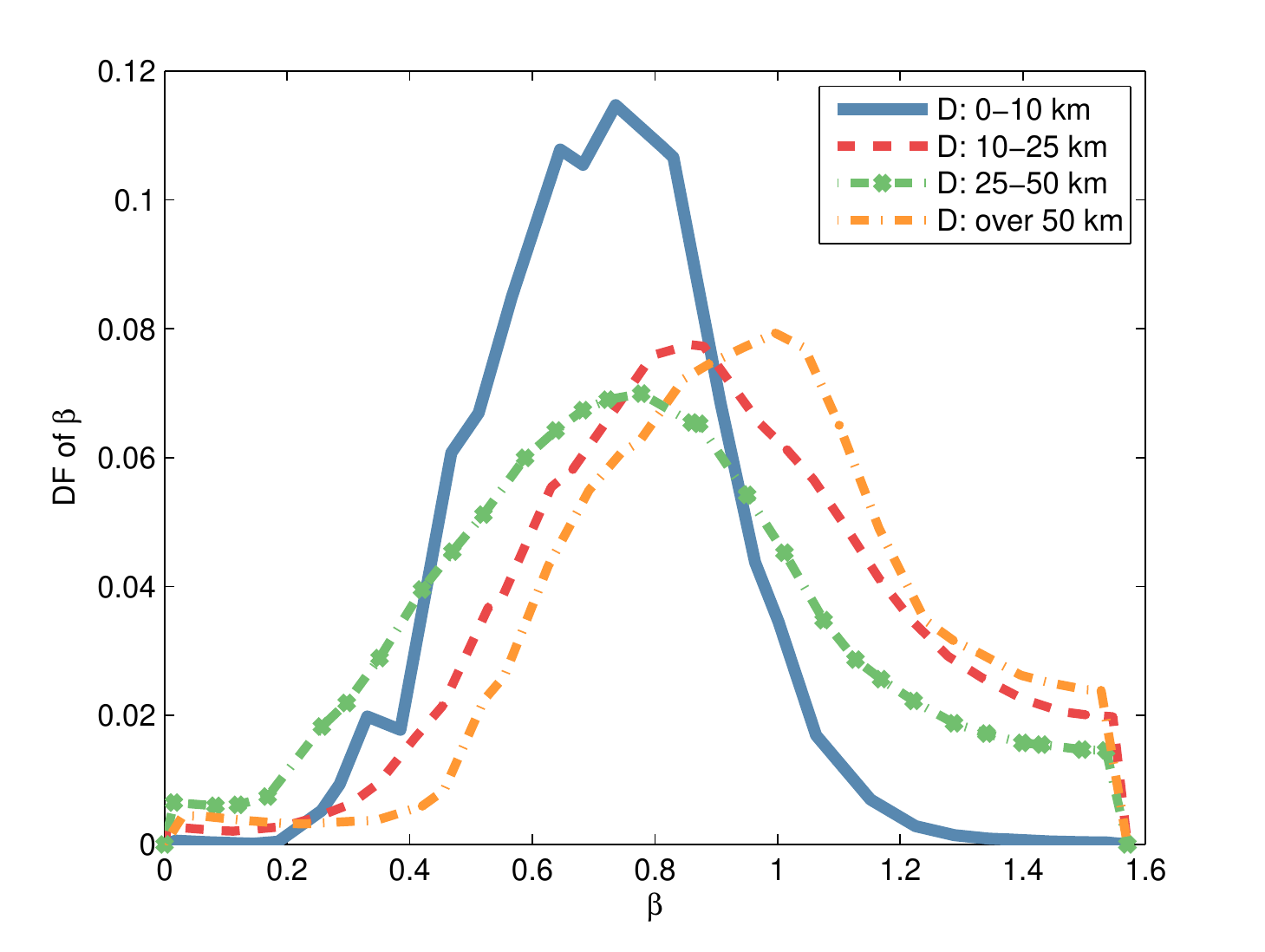} \caption{Comparison of the number densities of the spin $\beta$ for asteroids of different sizes from WISE data. The absolute value of the ecliptic latitude of the spin axis decreases from left (perpendicular to the ecliptic plane) to right (in the ecliptic plane).} \label{fig:WISE-sizecomp02} \end{figure}

Similarly, large asteroids tend to have their spin axes closer to the ecliptic plane than the small ones (see Fig.\ \ref{fig:WISE-sizecomp02}), which qualitatively agrees with the results of \cite{Hanus_et_al_2011} and \cite{Durech_et_al_2016}, who studied the distribution of spins of asteroids on a sample of several hundred individual models. Contrary to the results based on individual models where the spins are clustered towards poles of ecliptic ($\beta = 0$ in our notation), our analysis shows that small asteroids have a remarkably sharp peak extending some 15$^\circ$ on both sides of the ecliptic spin latitude of 50$^\circ$, but the lack of values $\beta \approx 0$ might be caused by some systematic effects of our simple model (but we note the effect of $\sin \beta$ in the plot as discussed in Sect 2.1).

When comparing asteroid families, the size of the targets may thus be a factor. The shape and/or spin difference between two families may be partly driven by the difference in their size distributions. On the other hand, the best statistical material is acquired without adding the size as another dimension in the distribution function since there usually are just not enough targets to split a family into size bins. Thus we report the family $p$ and $\beta$ distributions here as such, using all family members without considering the size distribution a bias factor, even though a closer analysis between families may require taking at least the division into small and intermediate sizes into account. The proportion of large asteroids is small, so their contribution to the population distributions is usually small as well. As an example, we investigated the Eos family of well over 3000 $\eta$ estimates.
We performed some comparisons of $p$ and $\beta$ distributions for the whole family and its subset of smaller bodies with a diameter less than 20 km. For the $p$ distribution, the inclusion of large asteroids mainly affected the width of the peak and there was no noticeable difference in the distributions. The differences were even smaller for the $\beta$ distribution. These results suggest that the biases caused by large objects are insignificant and we will include full populations in our examples.
For this and other comparison purposes, we propose a measure of difference tailored to our case (instead of the Kolmogorov--Smirnov test that typically produces indecisive statistics).

Let $S_1$ and $S_2$ be two sets of population samples. For $S_i$, let $F_p(S_i)$ and $F_{\beta}(S_i)$ be the CDFs for the marginal distributions of DF solutions for $p$ and $\beta$, respectively (normalized to the interval $[0,1]$). We define the statistical difference measure between $S_1$ and $S_2$ as
\begin{equation}
\left\{
\begin{aligned}
D_p(S_1, S_2) & = \alpha_k \norm{ F_p(S_1) - F_p(S_2) }_{k} \\
D_{\beta}(S_1, S_2) & = \alpha_k \norm{ F_{\beta}(S_1) - F_{\beta}(S_2) }_{k}
\end{aligned} \quad ,
\right.
\label{eq:popul-diff}
\end{equation}
where usually $k=1$, $k=2$ or $k=\infty$, and $\alpha_k$ is a norm-based scaling factor to fix the statistical difference to the same magnitude for all norms; typically, $\alpha_1 = 1/4$, $\alpha_2 = 1$ and $\alpha_{\infty} = 2$. The case $k=\infty$ is used in the Kolmogorov--Smirnov test. In addition to the $L^{\infty}$ norm, we will also compute the $L^1$ and $L^2$ norms, as this way we will have a better understanding of the type of statistical difference, for example, do the distributions differ in terms of the maximum or mean difference. Generally, our simulations suggest that in this context two distributions can be considered statistically different if $D \gtrsim 0.2$, although one number does not tell the whole story, and it is more instructive to perform a visual inspection on the marginal DF and CDF plots. 

\begin{figure}[!ht] \centering \includegraphics[width=0.242\textwidth]{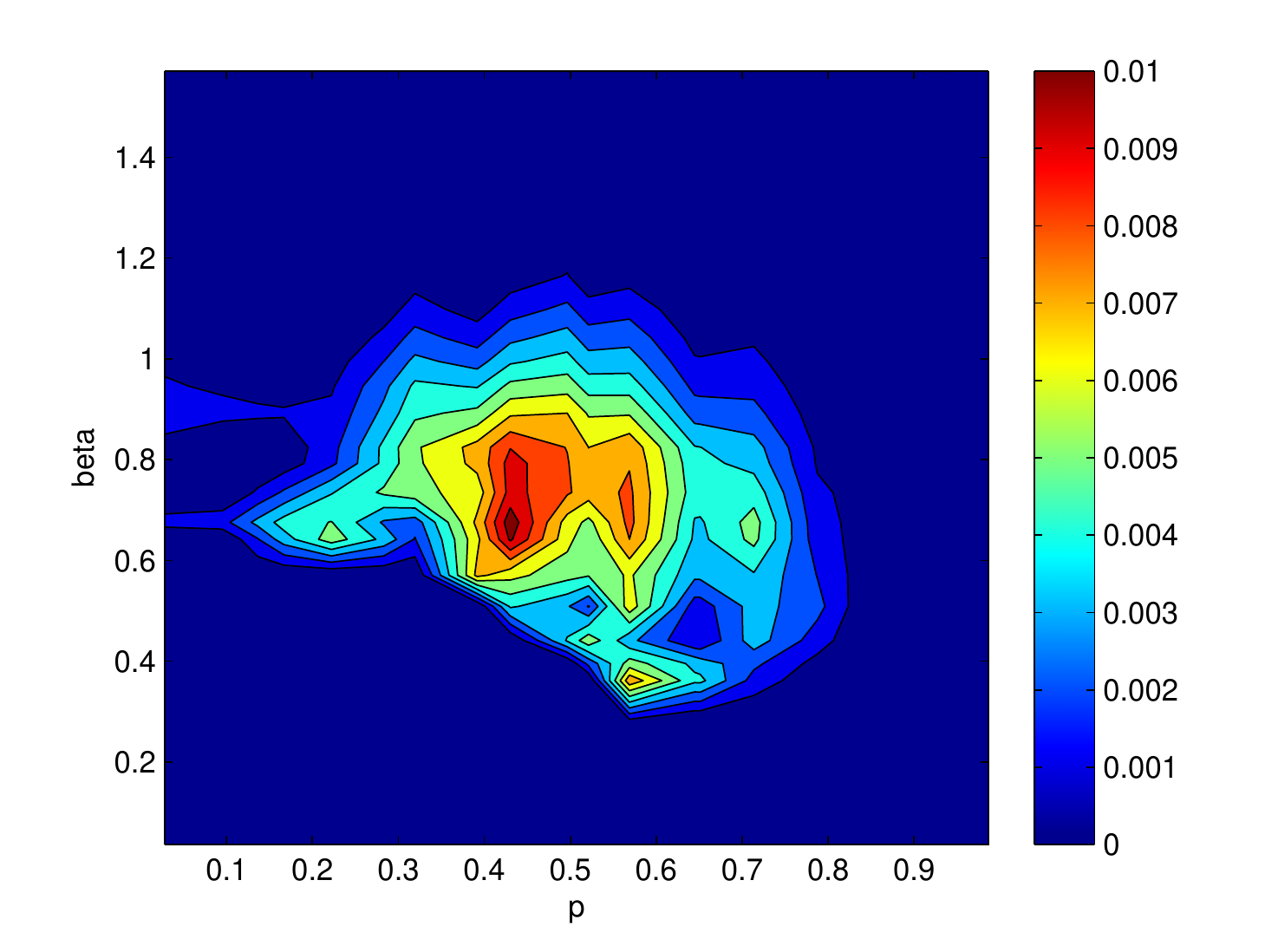} \includegraphics[width=0.242\textwidth]{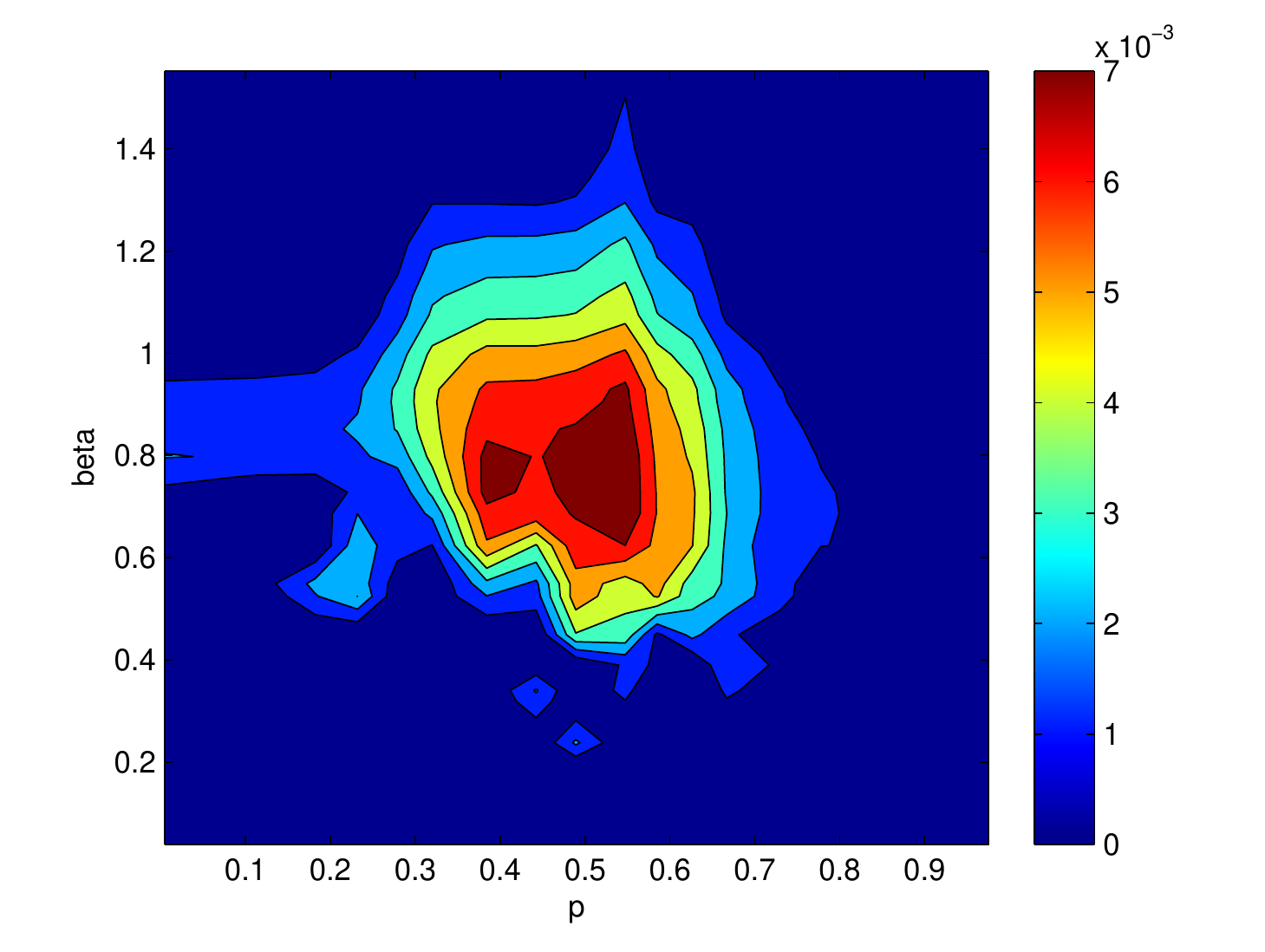}\caption{Contour solution of the joint $(p, \beta)$ distribution based on WISE (left) and WISE+ALCDEF data (right).  Deconvolution is not used to preserve the double-peak information.} \label{fig:WISEMPC} \end{figure}

\begin{figure}[!ht] \centering \includegraphics[width=0.49\textwidth]{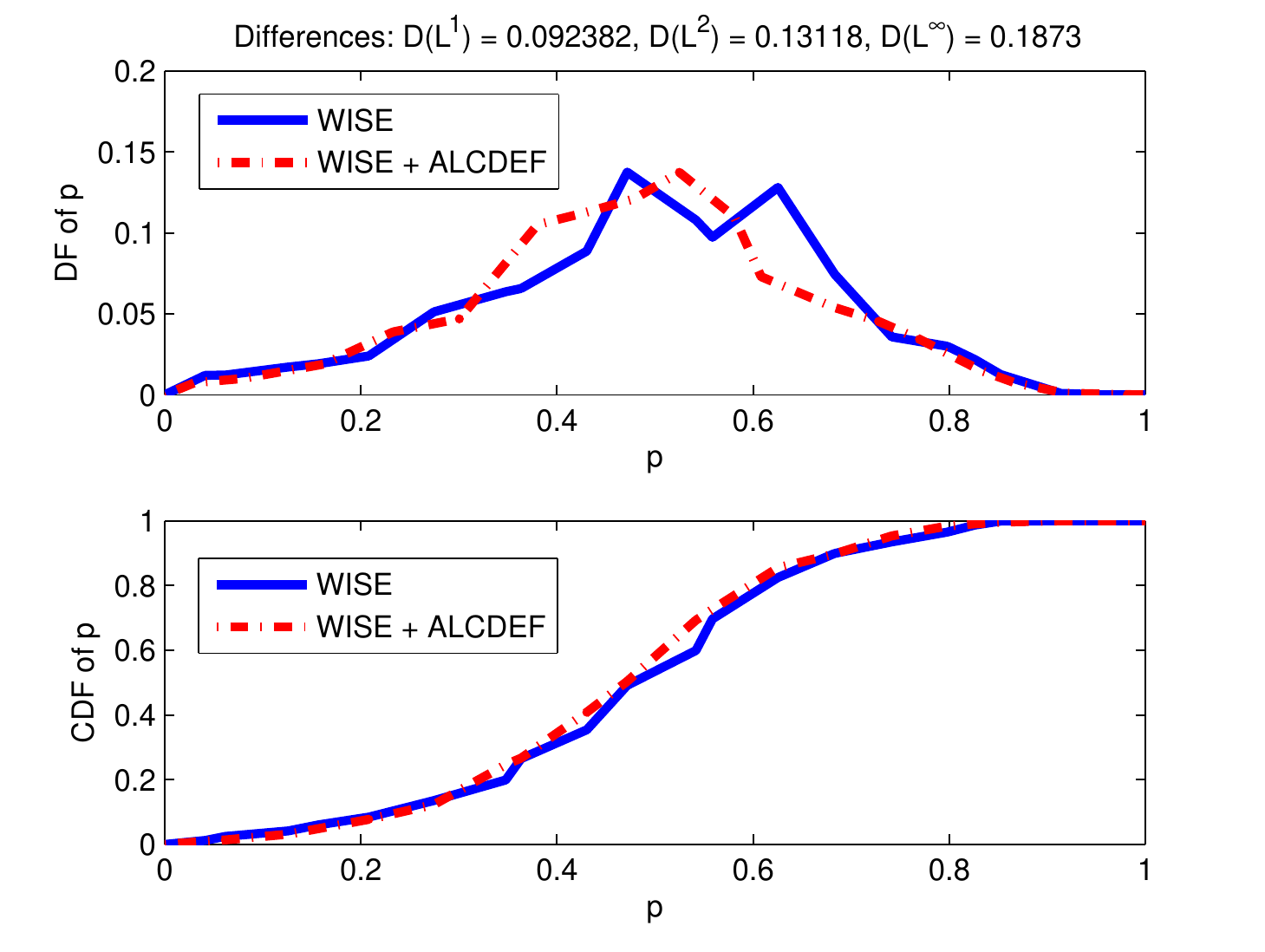} \caption{Comparison of the marginal DFs (top) of the shape elongation $p$ for the entire WISE population and merged $\eta$s from WISE and ALCDEF databases, and of their marginal CDFs (bottom).} \label{fig:WISE-MPC01} \end{figure}

\begin{figure}[!ht] \centering \includegraphics[width=0.49\textwidth]{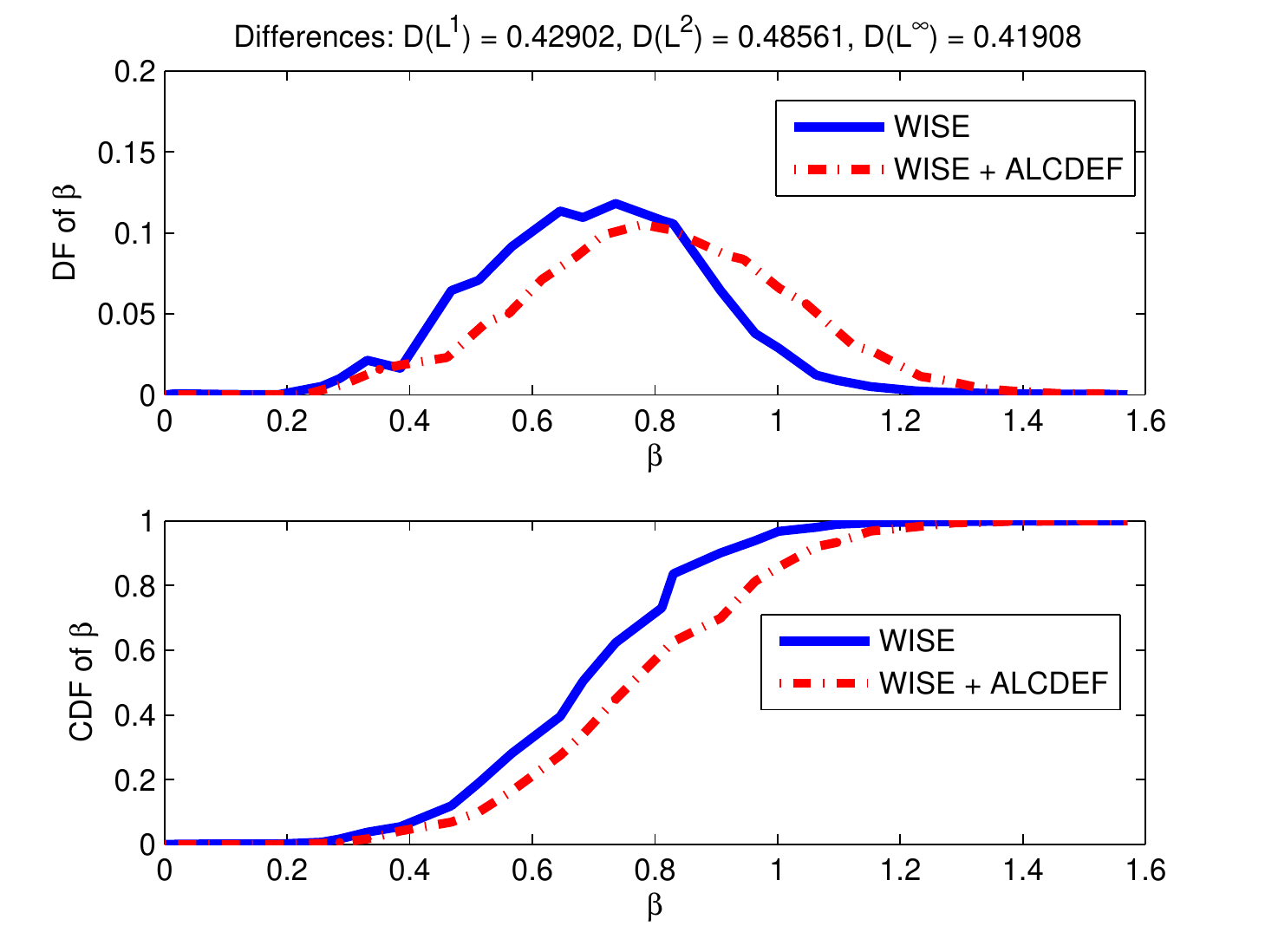} \caption{Comparison of the marginal DFs (top) of the spin $\beta$ for the entire WISE population and merged $\eta$s from WISE and ALCDEF databases, and of their marginal CDFs (bottom).} \label{fig:WISE-MPC02} \end{figure}

In Eq.\ \eqref{eq:popul-diff}, we chose to compare the CDFs rather than the DFs since the latter is the derivative of the former and CDFs are monotone functions, so computing the norm of their differences is more stable when one aims at one number depicting the difference. The CDF difference tells whether the distributions are really different in the first place, and the DFs give additional details of the potential differences.       

Finally, we ran tests to see if there is a bias associated with the selected database. WISE is one of the largest databases, with our method being able to cover about 85,000 asteroids. We can get about 86,000 values of $\eta$ from WISE, which means we can get approximately one value for our observable from each asteroid. Indeed, WISE is one of the biggest asteroid databases available. The average number of brightness measurements available for one $\eta$ is $\langle n_L \rangle \approx 9$. The ALCDEF lightcurve database contains about $14,000$ asteroids suitable for our method, and yields about $39,000$ values of $\eta$, resulting in less than three values of $\eta$ from each asteroid. For ALCDEF, $\langle n_L \rangle \approx 18$. Despite its smaller sample of asteroids, the ALCDEF's $\eta$-per-asteroid ratio is better than WISE's. The drawback of the ALCDEF is the weak or nonexistent calibration as well as selection effects. 

        \begin{figure}[!ht] \centering \includegraphics[width=0.49\textwidth]{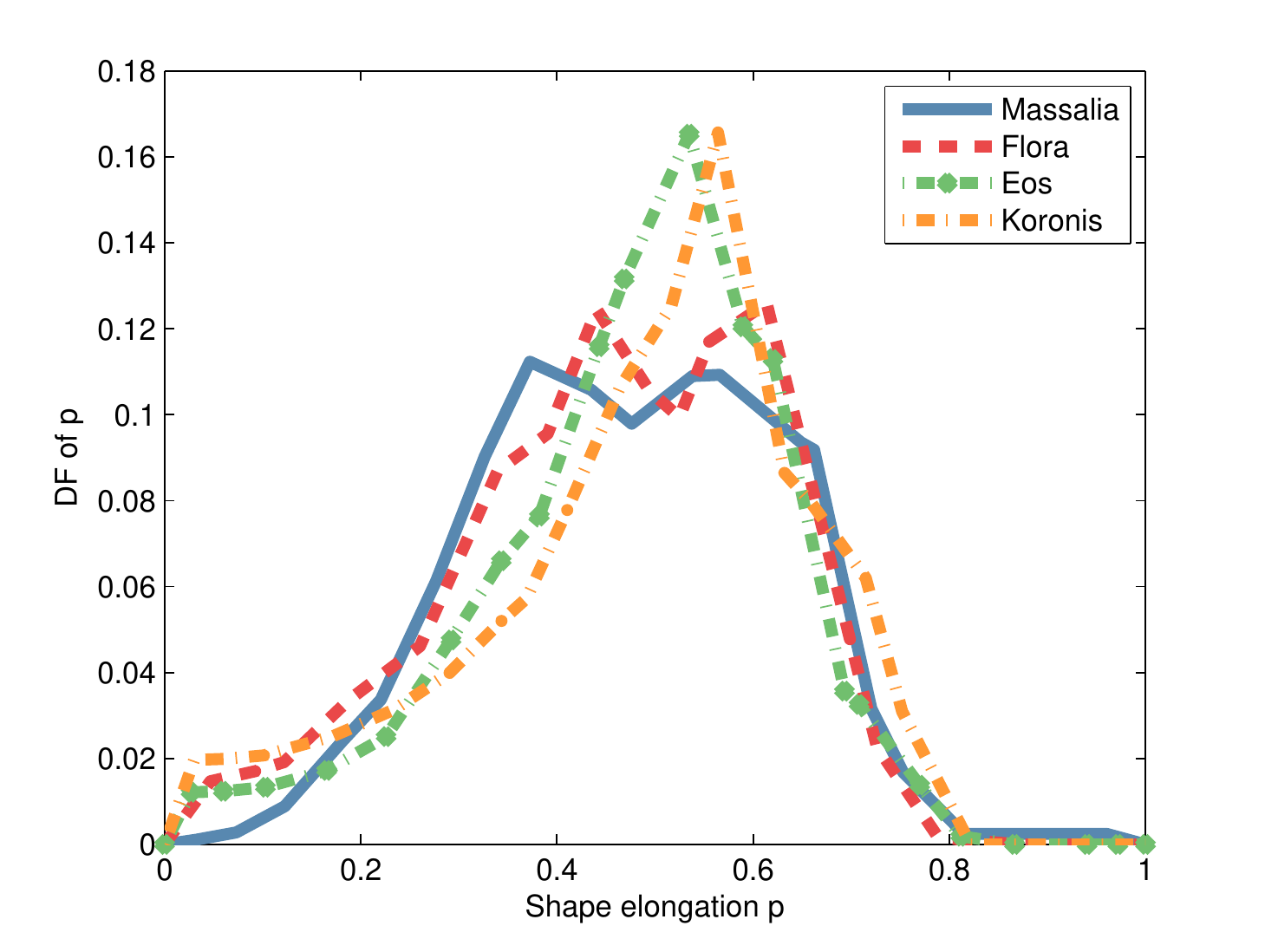} \caption{Obtained $p$ distributions for the asteroid families Massalia, Flora, Eos, and Koronis.} \label{fig:ast-p01} \end{figure}
        
        \begin{figure}[!ht] \centering \includegraphics[width=0.49\textwidth]{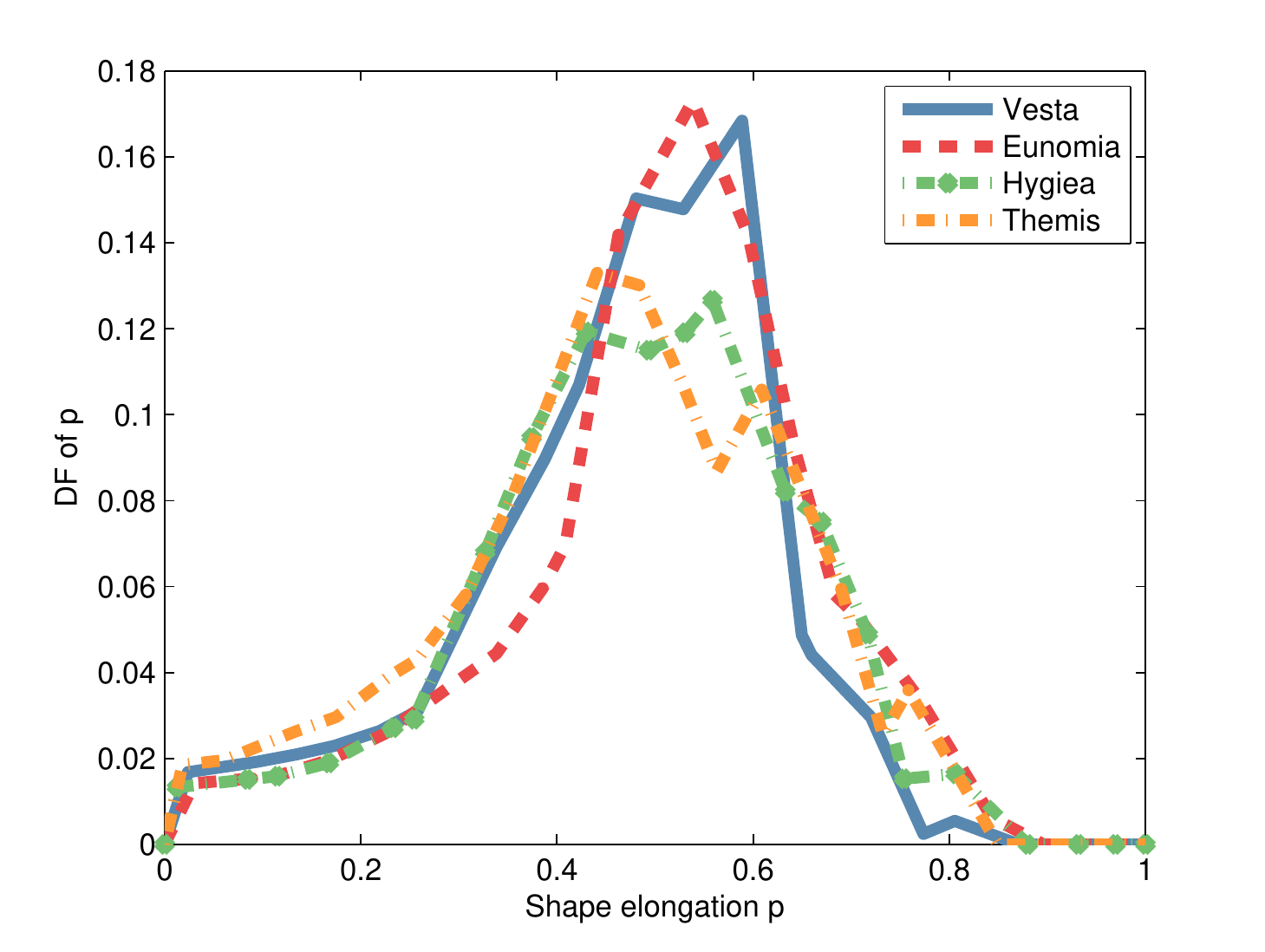} \caption{Obtained $p$ distributions for the asteroid families Vesta, Eunomia, Hygiea, and Themis.} \label{fig:ast-p02} \end{figure}

We perform a consistency check by comparing two distributions. First, we compute the solution of the inverse problem using all $\eta$s obtained from the WISE population. Then we do the same, but using all the $\eta$s from both WISE and ALCDEF. The $(p,\beta)$-plane plots are shown in 
Fig.\ \ref{fig:WISEMPC}.
        
The $p$ distributions of WISE and WISE+ALCDEF are plotted in Fig.\ \ref{fig:WISE-MPC01}, while the $\beta$ distributions of WISE and WISE+ALCDEF are plotted in Fig.\ \ref{fig:WISE-MPC02}. We have not used deconvolution procedures since, being designed for one-peak distributions, they would smooth out the two-peaked result. For $p$ distributions, a visual inspection shows some differences between the DFs, such as the unrealistic boost of $p$-values around 0.4 by the ALCDEF addition. The CDFs are similar enough to suggest that the added ALCDEF data do not greatly distort the $p$ distribution. For $\beta$, on the other hand, the addition of the ALCDEF data shifts the distribution to the right, and the $\beta$ distribution obtained from the hybrid data is obviously different from the one with WISE data only. Therefore, we conclude that there may be a database-related bias included, especially with the $\beta$ solutions, and it is advised to use caution in the selection of a database. Indeed, the ALCDEF data are distorted by a number of selection effects due to the visibility and popularity of the targets. WISE targets are more evenly and comprehensively spread and observed from a satellite, so the biases are smaller. Due to this and the sufficiently large number of database targets, we consider WISE data more reliable for distribution analysis and use them in the studies below.

        \begin{figure}[!ht] \centering \includegraphics[width=0.49\textwidth]{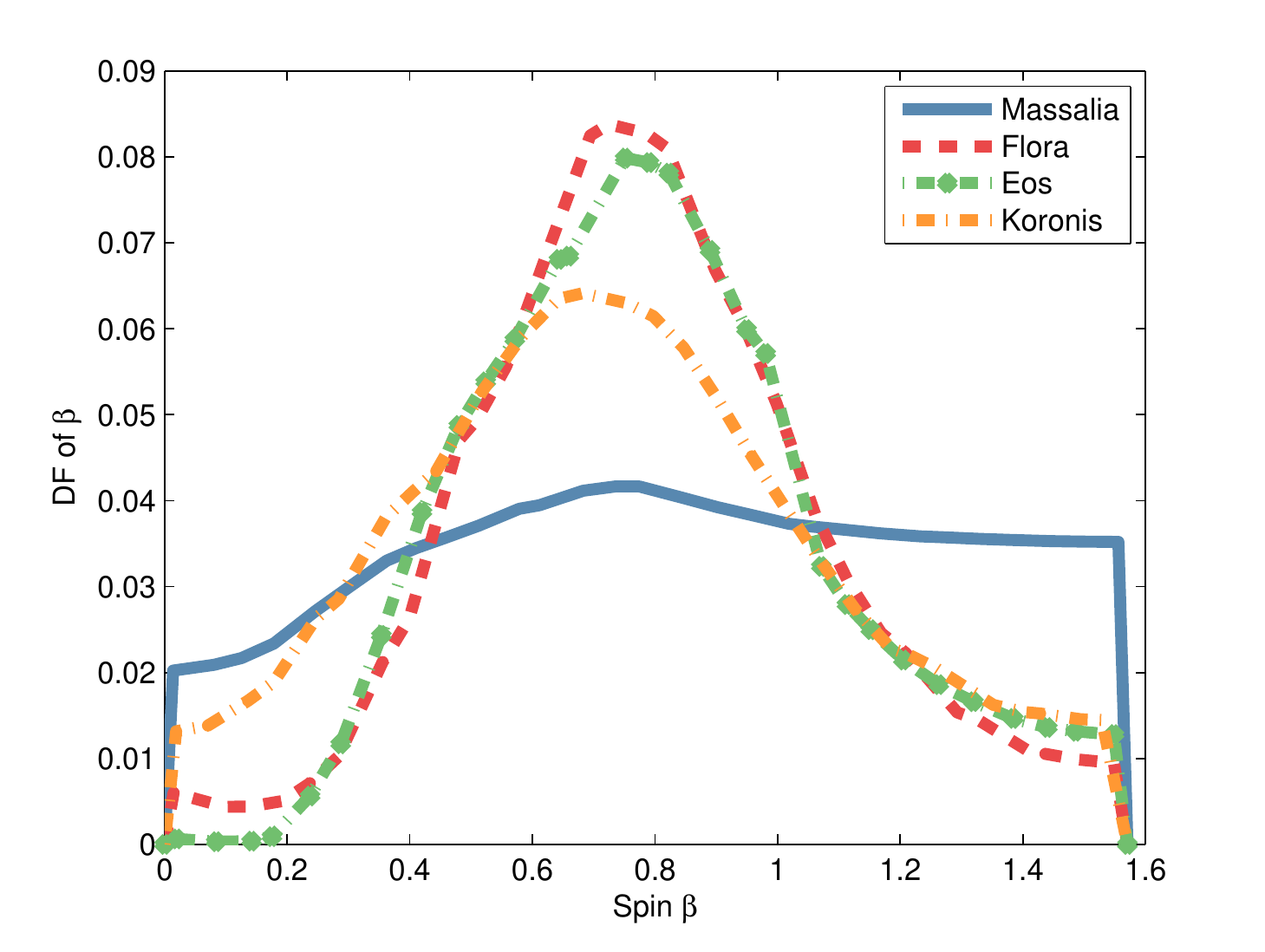} \caption{Obtained $\beta$ distributions for the asteroid families Massalia, Flora, Eos, and Koronis.} \label{fig:ast-beta01} \end{figure}
        
        \begin{figure}[!ht] \centering \includegraphics[width=0.49\textwidth]{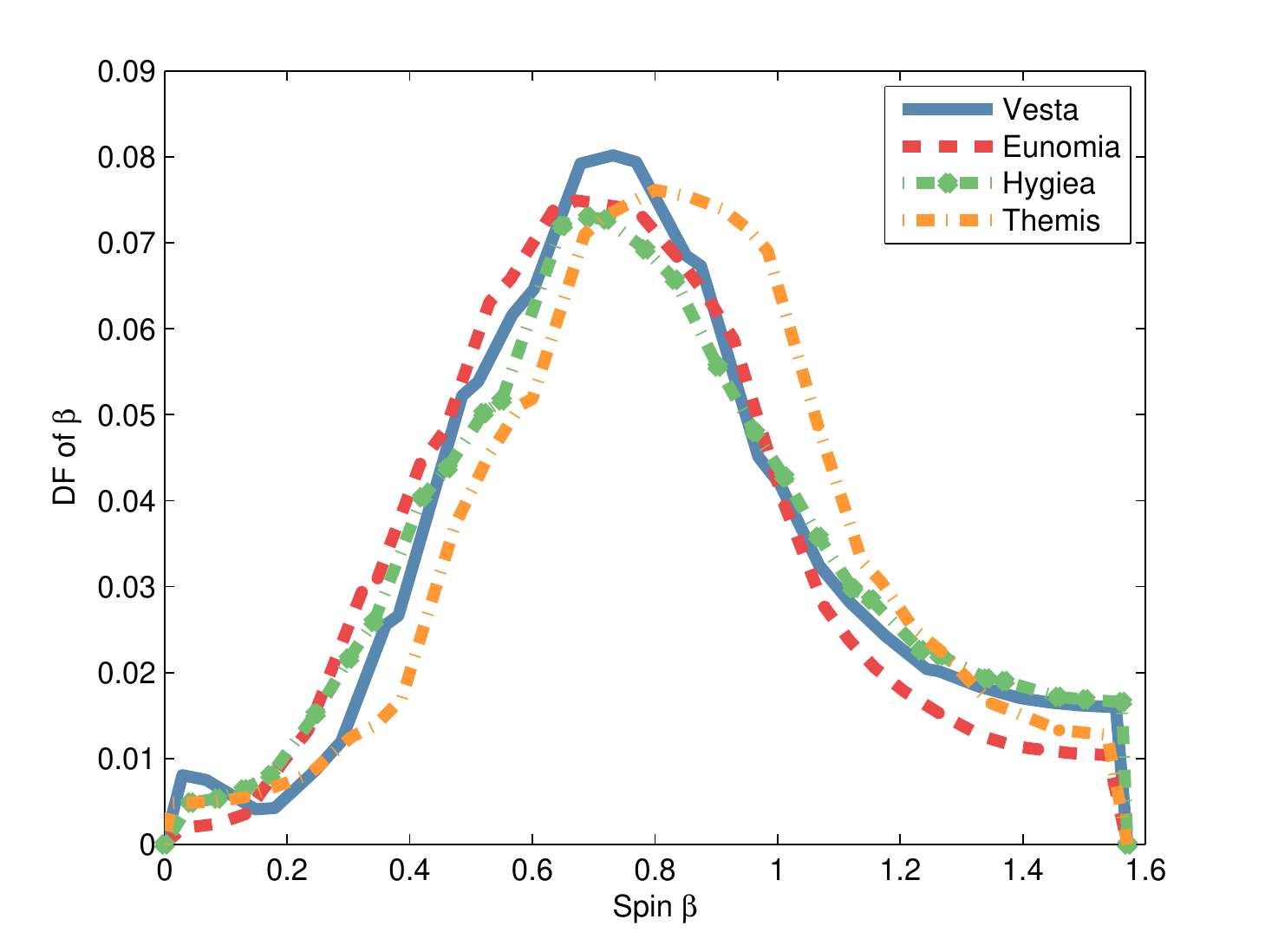} \caption{Obtained $\beta$ distributions for the asteroid families Vesta, Eunomia, Hygiea, and Themis.} \label{fig:ast-beta02} \end{figure}

\subsection{Examples of distributions and their comparison}

Below we list some results from family distribution reconstruction and comparison by their inferred marginal distributions of $p$ and $\beta$. We chose families that were interesting from a statistical point of view, mainly to demonstrate the features, differences, and similarities the method can discover. Obviously, in addition to the interpretation and analysis of the results, there are many other families as well as populations other than families to consider in further work from astronomical points of view. We note that the computed distributions for asteroid families vary slightly in the figures, just to illustrate that randomized inverse grids lead to slightly different details in distribution solutions. 
        
 The shape elongation and spin distributions for eight different asteroid families are shown in Figs.\ \ref{fig:ast-p01}--\ref{fig:ast-beta02}. The obtained shape elongation distributions are different from those obtained by \citet{szabo}, who assumed a uniform distribution of spin axes and utilized the less reliable two-point brightness scatter observable. The number of available asteroid samples per family in the WISE database varies. Flora and Eos have about 3,000 WISE samples, while Vesta, Eunomia, Hygiea, and Themis have 1,000--2,000. Koronis has 642 WISE samples, while Massalia is limited to only 154 samples. According to our simulations, solving the inverse problem several times for Koronis leads to fairly good regular solutions, so the sample size can be trusted to be large enough. Massalia, on the other hand, has stability problems with the small sample size, so its results cannot be considered to be as reliable as the others, but we include it for completeness. Typically, a sample of at least 500 objects is required in order to obtain stable solutions that can successfully recover from the model errors and noise.

\begin{figure}[!ht] \centering \includegraphics[width=0.242\textwidth]{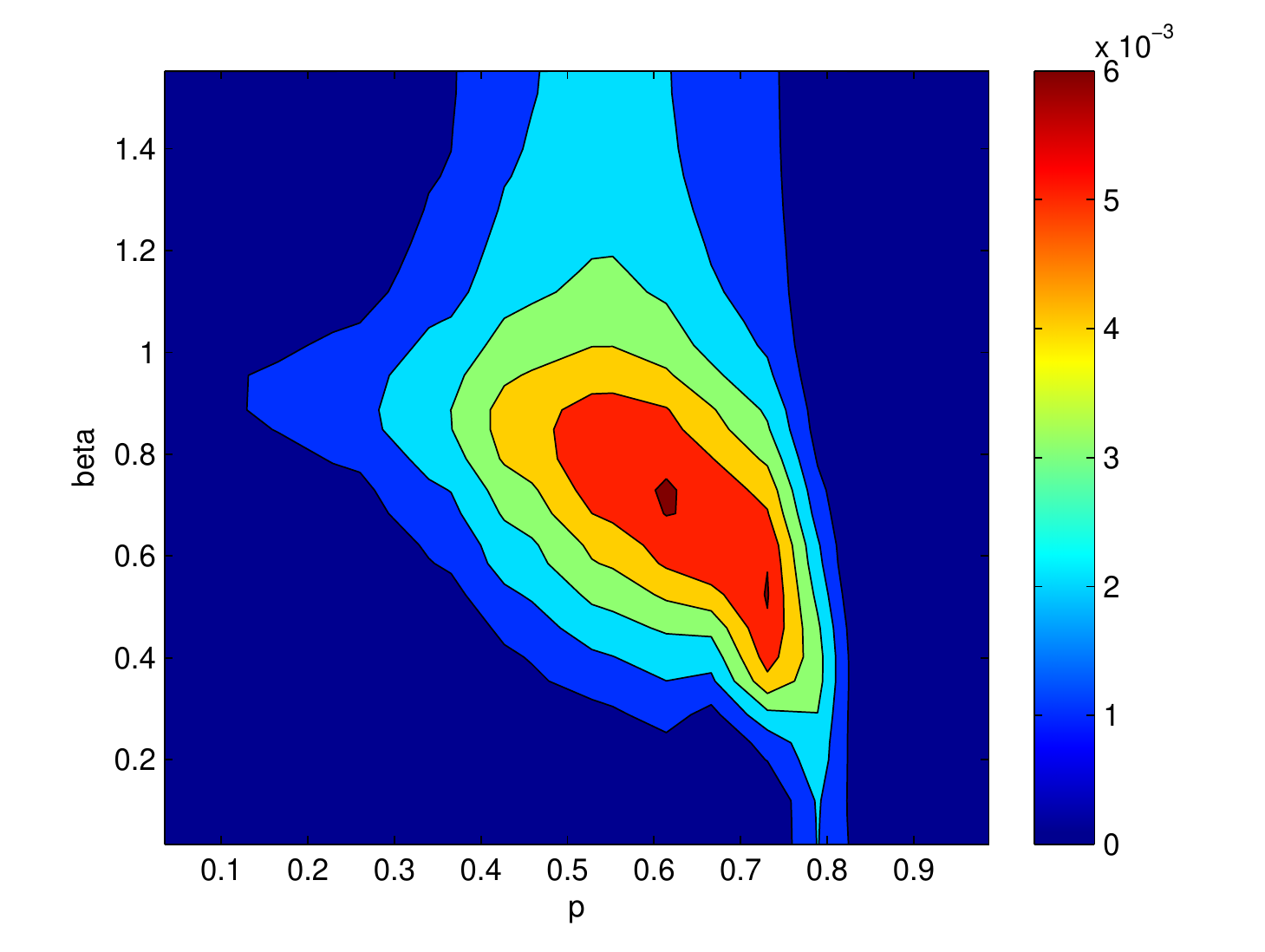} \includegraphics[width=0.242\textwidth]{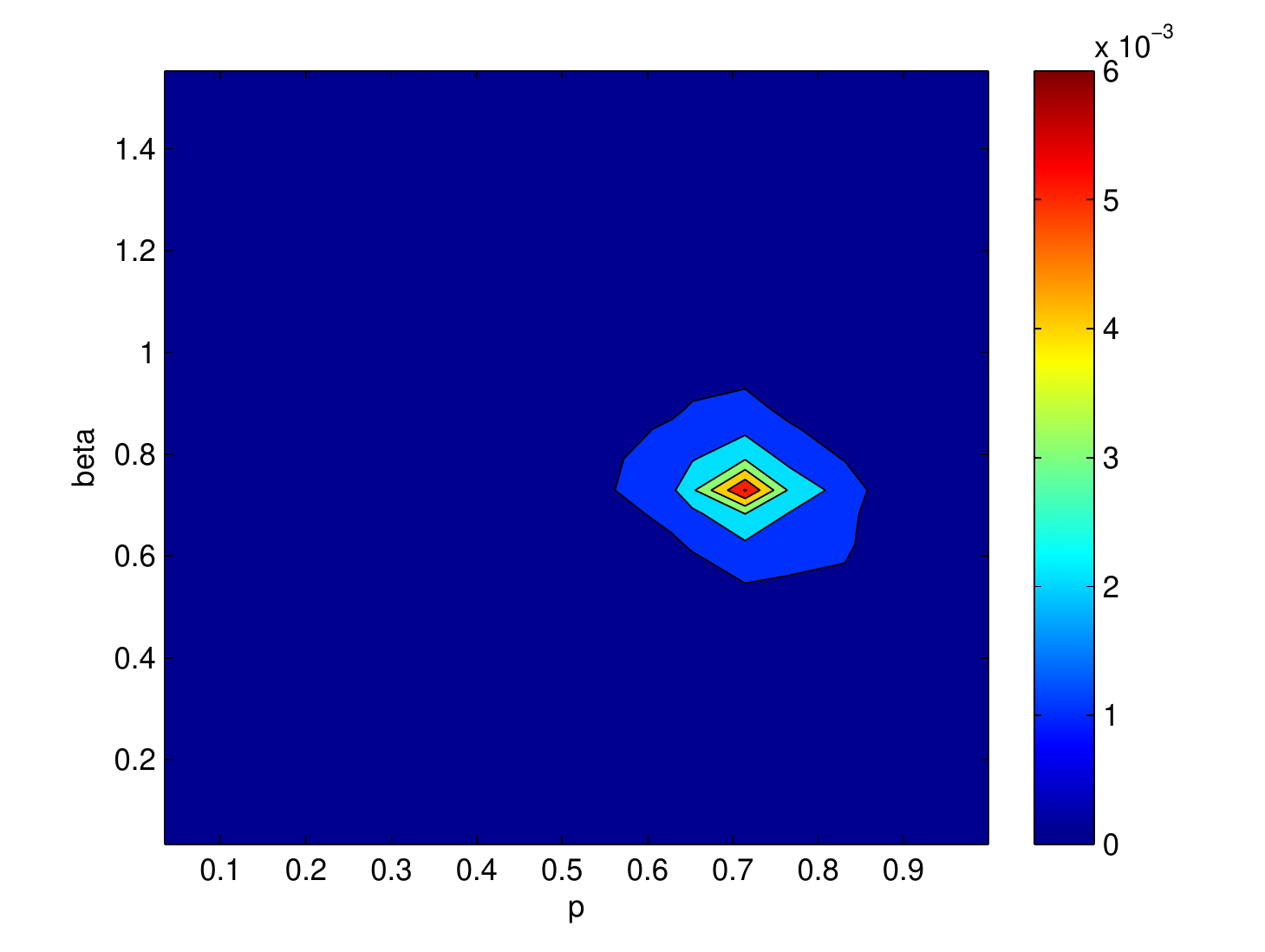} \includegraphics[width=0.49\textwidth]{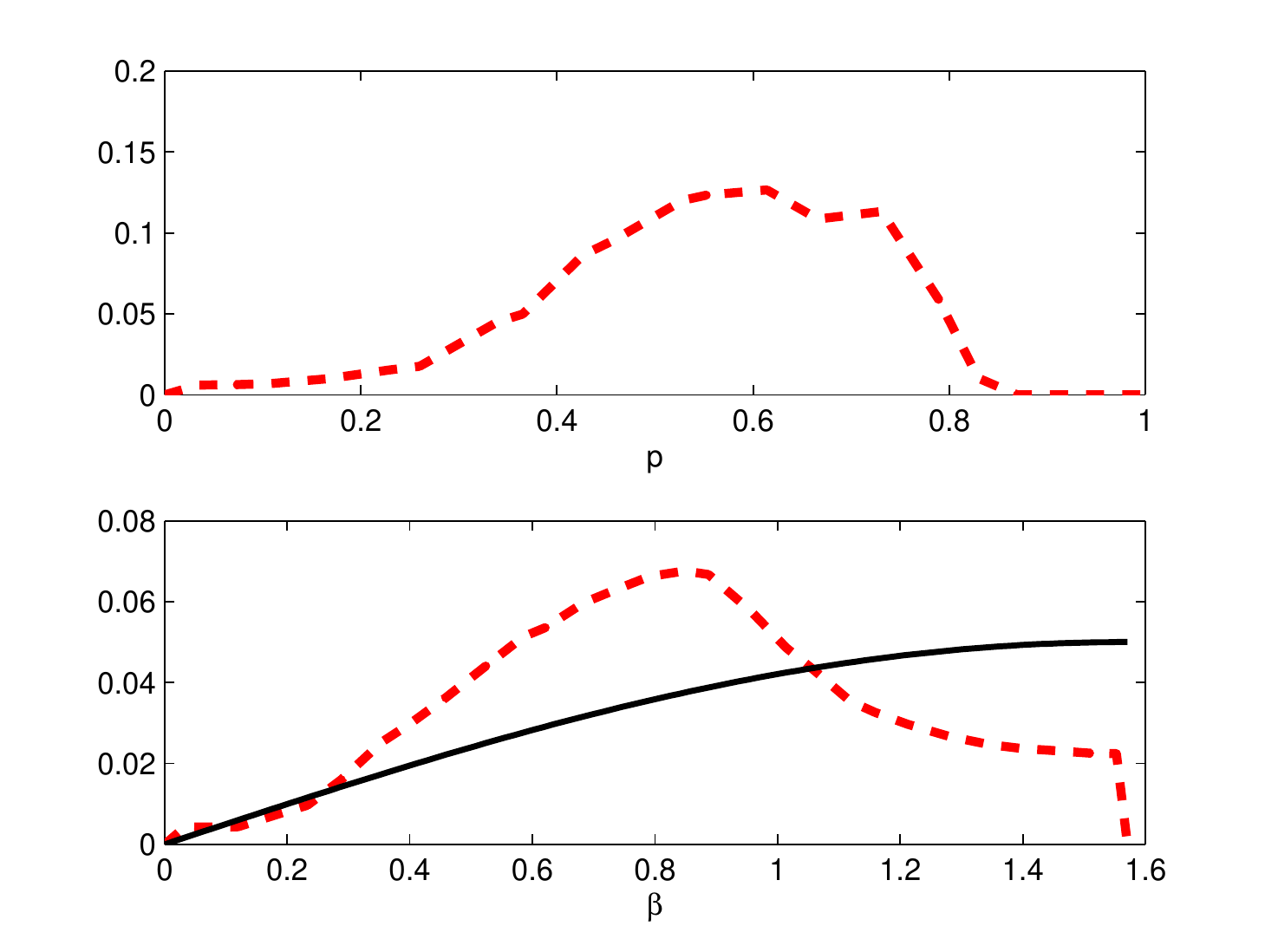}\caption{Contour solution of the joint $(p, \beta)$ distribution of the Alauda family (top left), the deconvoluted smoothing of the same solution (top right), and the normalized marginal distributions for $p$ (middle) and $\beta$ (bottom). The black solid curve ($\sin \beta$) depicts the curve shape of a constant level of spin distribution on the sphere. Deconvolution is not used for the marginal distributions in order to avoid the loss of information in the smoothing.} \label{fig:Alauda-all} \end{figure}

% Why Alauda: Alauda has many spherical asteroids               
We observed that the Alauda family (some 800 samples) contains a somewhat higher ratio of near-spherical bodies than other families. This is likely to be an intrinsic quality of the family, as the same result holds when the large asteroids ($D > 20$ km) have been filtered out. The contour solution of the joint $(p, \beta)$ distribution, both with and without deconvolution, as well as the marginal $p$ and $\beta$ distributions (without deconvolution) for the Alauda family are plotted in Fig.\ \ref{fig:Alauda-all}. We plotted an additional $\sin \beta$ curve in the $\beta$ plot to illustrate what the spin DF would look like if it was uniformly distributed on the sphere.

                        \begin{figure}[!ht] \centering \includegraphics[width=0.49\textwidth]{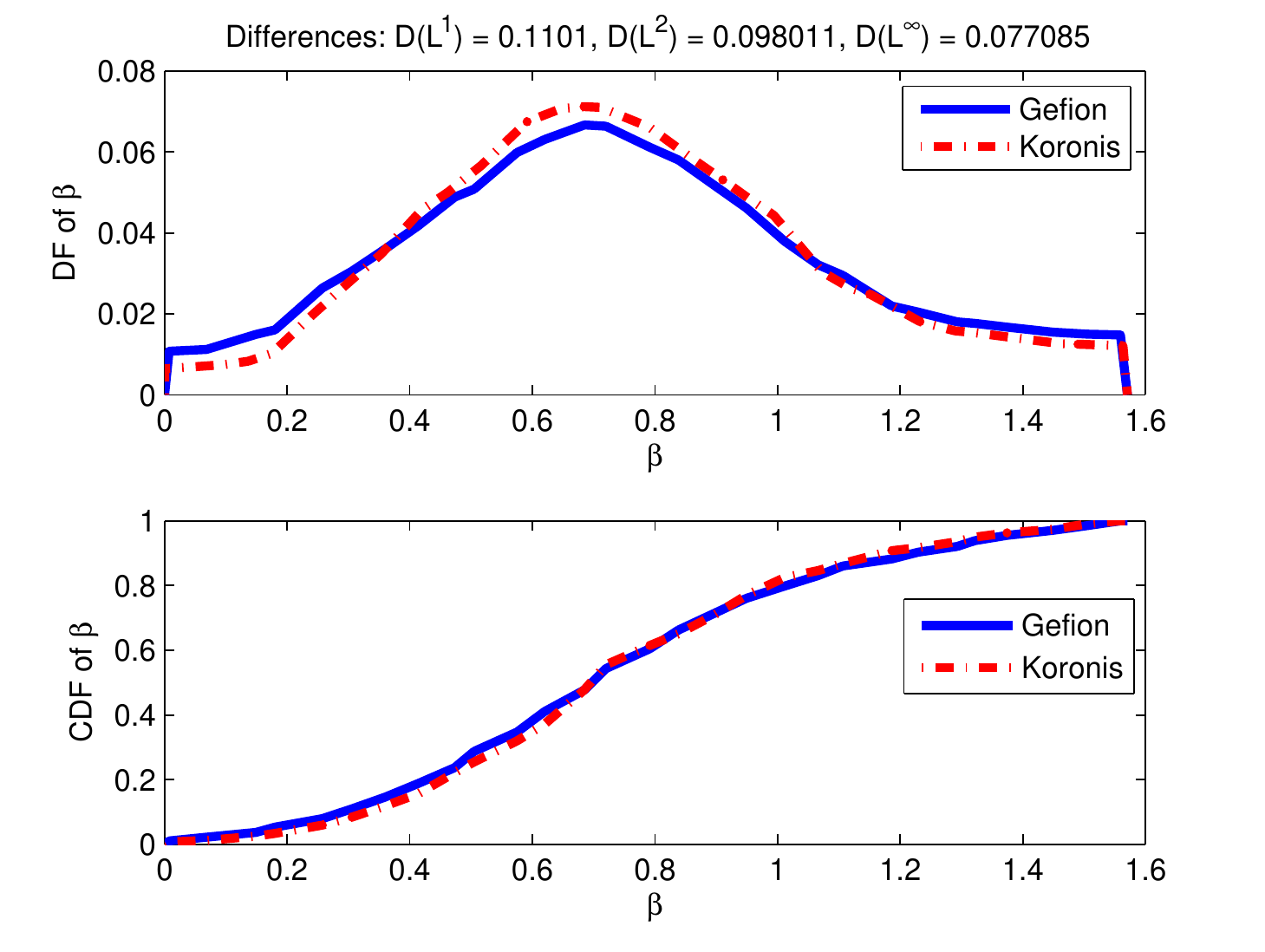} \caption{Comparison of the marginal DFs (top) of the spin $\beta$ for Gefion and Koronis families, and of their marginal CDFs (bottom). } \label{fig:GefionKoronisB} \end{figure}

\begin{figure}[!ht] \centering \includegraphics[width=0.49\textwidth]{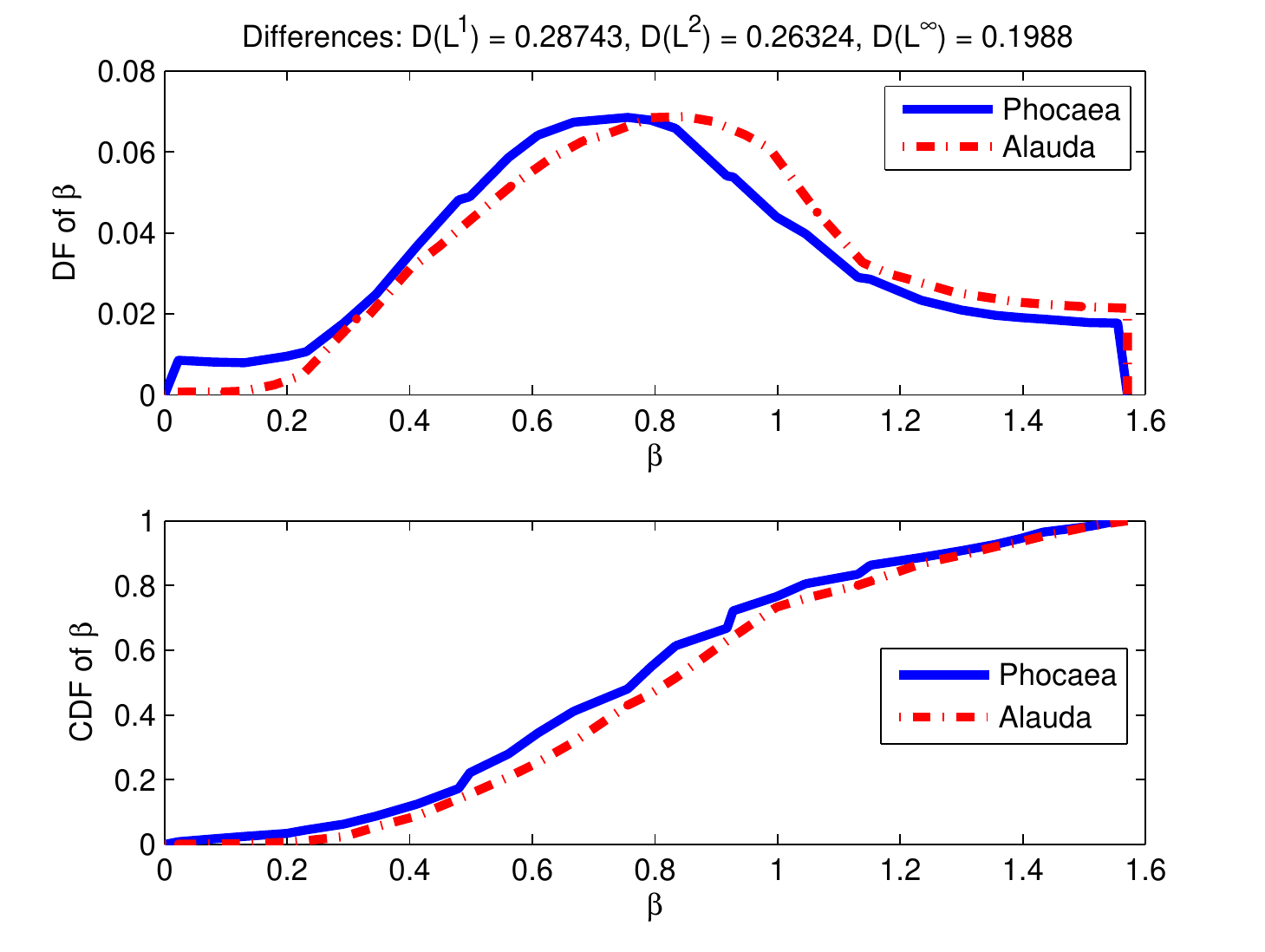} \caption{Comparison of the marginal DFs (top) of the spin $\beta$ for Phocaea and Alauda families, and of their marginal CDFs (bottom). } \label{fig:PhocaeaAlauda} \end{figure}

\begin{figure}[!ht] \centering \includegraphics[width=0.49\textwidth]{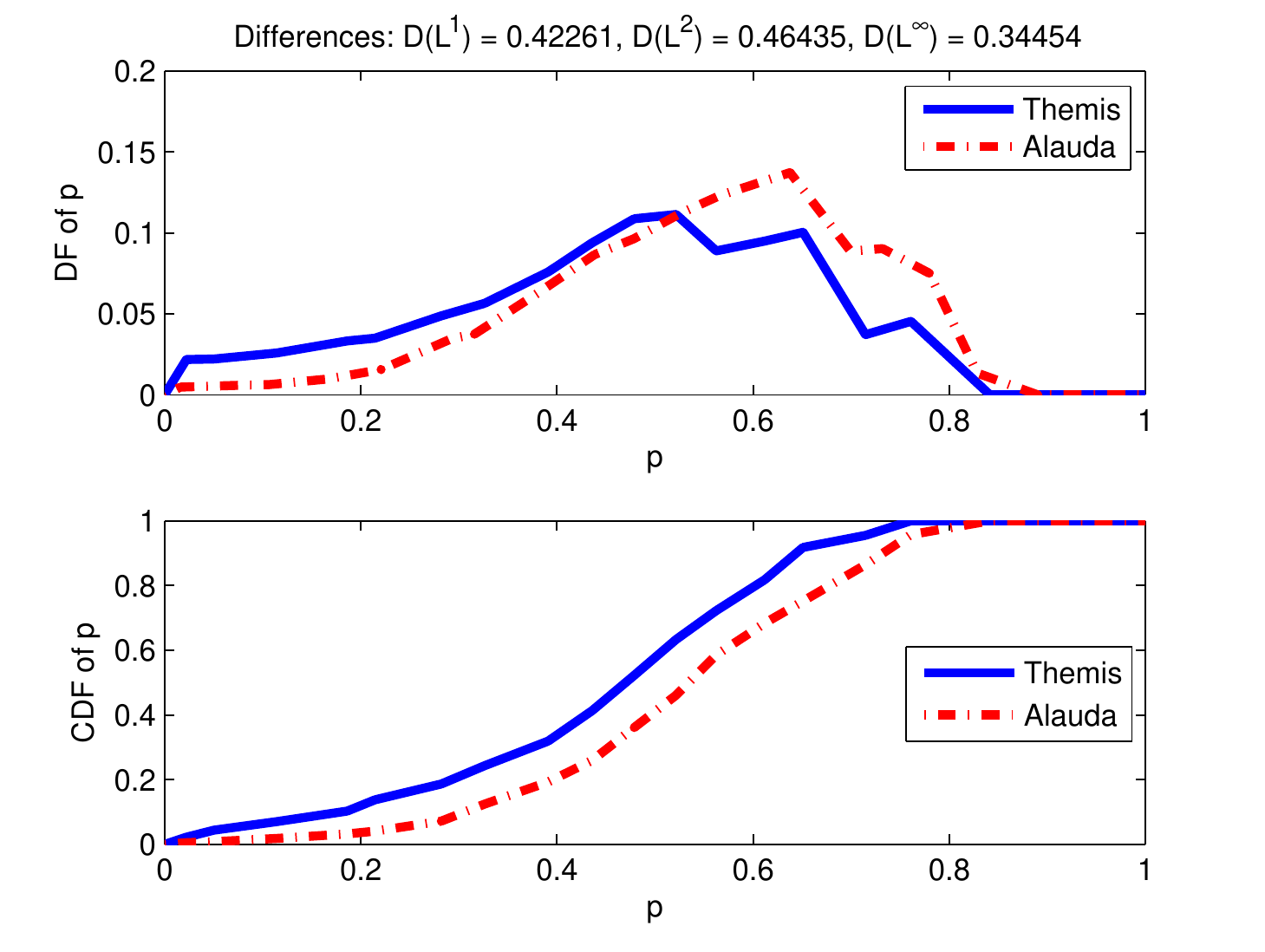} \caption{Comparison of the marginal DFs (top) of the shape elongation $p$ for Themis and Alauda families, and of their marginal CDFs (bottom).} \label{fig:ThemisAlauda} \end{figure}

To show examples of difference  classes between families, we consider cases of small, borderline, and large values of the difference measures.
The very similar $\beta$ distributions of the Gefion and Koronis families are shown in Fig.\ \ref{fig:GefionKoronisB}, while
the $\beta$ distributions of the Phocaea (some 1000 samples) and Alauda families are plotted in Fig.\ \ref{fig:PhocaeaAlauda}. Phocaea has slightly more asteroids with spins closer to perpendicular to the ecliptic plane, while both families have heavy tails close to the plane. Generally, our solutions appear to give less weight to the values of $\beta$ close to $0$ or $\pi/2$. In the case of $\beta=0$, this is partly due to the factor of $\sin\beta$ in the occupation numbers. For $\beta=\pi/2$, this can be due to, for example, size distribution, noise, and orbit positions away from the ecliptic plane. In our simulations, we did not find any particular mechanism or tendency for the scarcity of solutions close to the ecliptic plane. Finally, in Fig.\ \ref{fig:ThemisAlauda}, we give an example of the clearly different shape distributions of the Themis and Alauda families.

To confirm the reliability of distributions of $\beta$ and $p$ for individual families, it would be ideal to compare our CDFs with those constructed from individual models derived by lightcurve inversion. Unfortunately, this is not possible at this stage, because the number of known models for a typical family is a few tens at most \citep{Hanus_et_al_2013}. Another possibility would be to compare CDFs reconstructed from different and independent data sets. For example, the difference between Themis and Alauda families is also significant when we do the same analysis with the Panoramic Survey Telescope and Rapid Response System (Pan-STARRS) data (Cibulkov\'a et al., in prep.).

\section{Discussion and conclusions}

The statistical CDF approach is a fast way of testing hypotheses about shape and spin distributions of asteroid populations without constructing models of separate objects. It is applicable for discovering the existence of peaks in distributions of parameters and for comparing the distributions of different populations. 

There are numerous possibilities of analysing and comparing asteroid families, and a comprehensive analysis and interpretation of results is not the aim of this paper. Our main goal was to discuss the usefulness and various aspects of the CDF approach, and build mathematical tools for its efficient use. As we have seen, the data and inversion procedures of the problem are, in fact, very simple and fast to generate and apply as such. The main burden lies in the judicious interpretation of the results. A scrutiny of the usable databases may well be necessary, as some data sources may cause skewed results due to biases and/or noise.

We introduced a robust observable that can provide information on both the shape elongation and spin properties of asteroid populations, and performed an analysis on the theoretical background, providing also some examples of the obtained distributions, inspected from the statistical point of view. In our analysis, we proved that unique solutions can be obtained for both shape elongation and spin distributions, and we performed numerical simulations in order to verify that they coincide with the analytical results. It is interesting to note that, while the abundance of orbits close to the ecliptic plane means that many individual asteroid models necessarily have an ambiguity of 180 degrees in the spin longitude \citep{genproj}, the same ecliptic orbital configuration makes possible the population-level information on spin latitudes.

Due to the model noise and the assumptions made, we cannot expect to obtain detailed, high-resolution solutions of the distributions, but we get the overall picture when the observational noise is sufficiently low (at most 0.05 mag or so). If the data noise is large, the whole point of using the brightness variations as the observable is challenged, and the prior information needed for regularization would dominate the solution. Typically, high noise means that the real $p$-information on near-spherical bodies disappears, and the $\beta$-information is similarly severely diluted. For low observational noise, we can additionally use a deconvolution filter to correct the systematic errors caused by modelling errors and noise. The deconvolution is a visual tool which attempts to illustrate what the distribution of the parameters actually looks like, based on prior information obtained from simulations performed for synthetic data.

        Trans-Neptunian objects (TNOs) are especially interesting from the statistical point of view due to their slow orbital motion. Since their observing geometries change very slowly, proper individual models of TNOs cannot be made with the current ground-based instruments in less than tens of years. Also, for TNOs the solar phase angle is essentially zero, leading to ambiguous shape solutions from photometry. The statistical approach, however, applies to TNOs just as well as to other populations so long as there are sufficient targets in the observed set. This is, in fact, the only way to model TNO populations with data from large-scale surveys. A bonus with TNOs is that, because of the essentially fixed geometry and near-zero solar phase angle, all calibrated survey data points are usable for $\eta$-estimation even if they are separated by long time intervals. With main-belt asteroids, sparse data points from one apparition may be usable together for $\eta$ estimates if one uses a solar phase correction as in \citet{sparse}, \citet{sparseD}, or \citet{cib}. The additional systematic error from this is not necessarily very large considering the total error budget.

We emphasize that the statistical use of brightness variation, while a promising approach, should always be treated with caution. Above all, any proposed type of observable and implementation should be checked with realistic simulations where the synthetic data are created with a model different from the one used in inversion. The information potential of each dataset should be assessed by using its actual observing geometries in the simulations. These simulations yield insight into the uniqueness and stability properties and accuracy expectations. Based on analytical considerations and simulations mimicking real databases, we advocate the use of the $\eta$ observable and the corresponding analytical basis functions in the inverse problem.

We plan to offer a software application as a statistical (and simulation) tool that can be used for experimenting with different populations that are defined by the user. It may also be useful to construct solution procedures tailor-made for input populations. For example, one can create basis functions numerically by making synthetic CDFs for each $(p,\beta)$-bin with DAMIT-based shapes placed in the orbits of the populations.

\begin{acknowledgements}

We would like to thank Matti Viikinkoski for valuable comments and discussions as well as assistance with software. This research was supported by the Academy of Finland (Centre of Excellence in Inverse Problems), and HN was supported by the grant of Jenny and Antti Wihuri Foundation.
J\v{D} and HC were supported by the grant 15-04816S of the Czech Science Foundation. VAL has received funding from the European Union's Horizon 2020 research and innovation programme under grant agreements No. 640351 and No. 687378. This publication also makes use of data products from NEOWISE, which is a project of the Jet Propulsion Laboratory/California Institute of Technology, funded by the Planetary Science Division of the National Aeronautics and Space Administration. In addition, this research made use of the NASA/IPAC Infrared Science Archive, which is operated by the Jet Propulsion Laboratory/California Institute of Technology, under contract with the National Aeronautics and Space Administration.

\end{acknowledgements}

%-------------------------------------------------------------------

\begin{appendix} %First online appendix

\section{CDF integrals and analytical basis functions} \label{app:basis}

We recall the expression of the amplitude $A$, given by Eq.\ \eqref{amp}:
$$
A=\sqrt{1+(p^2-1)\sin^2\theta}.
$$
From this expression, the curves of constant $A$ in the $(p,\theta)$-plane are given by
\begin{equation}
\cos^2\theta_A(p)=\frac{A^2-p^2}{1-p^2}:=g_A(p). \label{gA}
\end{equation}
The solutions for $\theta_A$ are convex "ripples" starting from the point $(p=0,\theta=\pi/2)$
(upper left corner) for $A=0$ and continuing to the 
lines $\theta=0$ and $p=1$ for $A=1$ (lower right corner).
Denoting the model DF of elongation by $f(p)$, we write the unnormalized CDF $C(A)$ as 
$$
C(A)=\int_0^{p_{\rm max}(A)} f(p)\int_{\theta_A(p)}^{\pi/2} \sin\theta\,
d\theta \,dp,
$$
where the minimal shape elongation needed to produce
amplitude $A$, obtained at $\theta=\pi/2$, is
$p_{\rm max}(A)=A$.
With a change of variable $x = \cos \theta$, we get
\begin{equation}
C(A)=\int_0^A f(p)\int_0^{\sqrt{g_A(p)}} \,dx \,dp
=\int_0^A f(p) \sqrt{g_A(p)} \,dp.\label{CAp}
\end{equation}

We can also include the effect of spin distribution. Assuming $\lambda$ to
be isotropic and the observation directions to be in the $xy$-plane of the inertial frame (as they approximately are for the majority of asteroids, when this plane is that of the Earth's orbit), we study the DF $f_\beta(\beta)$ 
(or the joint DF $f(p,\beta)$ with $p$). The minimal aspect angle is
 $\theta_{\rm min}=\pi/2-\beta$. Now, substituting $ {\bf e}=(\cos\lambda_e,\sin\lambda_e,0)$ into
$\cos\theta=e_1\sin\beta\cos\lambda+e_2\sin\beta\sin\lambda+e_3\cos\beta$,
we have
$$
\cos\theta=\sin\beta\cos\Lambda,
$$
where $\Lambda:=\lambda-\lambda_e$ is assumed isotropic (evenly distributed 
longitudes of 
spins and observing directions). It is sufficient to explore the region
$\Lambda\in[0,\pi/2]$ as other quadrants are just symmetric multiples.

The curves of constant $\theta$ 
\begin{equation}
\Lambda_\theta (\beta)=\arccos\frac{\cos\theta}{\sin\beta}
\label{eq:isoth}
\end{equation}
in the $(\beta,\Lambda)$-plane are now expanding
``ripples'' of increasing $\theta$
starting from the point $(\beta=\pi/2,\Lambda=0)$ for $\theta=0$.
The CDF for $\theta$ is, with $x = \cos\beta$ (but retaining the argument $\beta$ in the DF for convenience),
\[
\begin{split}
C_\theta(\theta)&=\int_{\pi/2-\theta}^{\pi/2} 
f_\beta(\beta) \sin\beta\int_0^{\Lambda_\theta (\beta)} 
\,d\Lambda \,d\beta \\
&=\int_{\pi/2-\theta}^{\pi/2} f_\beta(\beta) \sin\beta
\Lambda_\theta (\beta)\,d\beta\\
&\stackrel{\text{Eq.\ } \eqref{eq:isoth}}{=}\int_{0}^{\sin\theta}
f_\beta(\beta)\arccos\frac{\cos\theta}{\sqrt{1-x^2}}\,dx.
\end{split}
\]
(Differentiating
$d C_\theta(\theta)/d\theta$ yields $\sin\theta$ when $f_\beta=1$ as expected for isotropic spins.)

Using the complement of $C_\theta$ (i.e. $\hat C_\theta$ in the decreasing direction from $\theta=\pi/2$ to $\theta=0$) to
write the number of  states between $\theta_A(p)$ and $\theta=\pi/2$, our CDF $C(A)$ is, analogously with Eq.\ \ref{CAp} and using
Eq.\ \ref{gA},
\begin{equation}
\begin{split}
C(A) = &\int_0^A \Big\lbrack\frac{\pi}{2}\int_0^1 f(p,\beta)\,dx \\ & - \int_{0}^{\sqrt{1-g_A(p)}}
f(p,\beta)\arccos\frac{\sqrt{g_A(p)}}{\sqrt{1-x^2}}\,dx\Big\rbrack \,dp.\label{CApx}
\end{split}
\end{equation}
(The use of $x$ is merely a matter of convenience for the integration limits.)

The basis functions, that is, any CDF $C(A)$ caused by all objects having given fixed $p_i$ and $\beta_j$, are now easy to write in closed form. They are obtained by replacing $p$ and
$\beta$ in the integrands by the fixed $p_i$ and $\beta_j$, setting $f=1$, and using the integration limits to describe the inequalities between $A$ and $p,\beta$ to define the piecewise function $C(A)$.  This replaces the integral by a sum of such basis functions each multiplied by the corresponding weight of the $p_i$ and $\beta_j$ bin. The resulting basis functions are given in Sect.\ 2.1.

The above assumption of most orbits to be close to the ecliptic plane is only approximate, and one can always define populations (especially those of near-Earth asteroids) for which it is not true even approximately. Thus the validity of this assumption should be checked for the targets used. However, as we show in Sect.\ 4, the assumption works quite well (given the large model error budget in any case) with typical asteroid populations for which there is some concentration of viewing geometries sufficiently near the ecliptic plane.

\section{Ill-posedness caused by two-point variation observables} \label{app:2point}

As earlier, we consider the case when $\theta$ is (approximately) the same for the pair. Now we have,
for two rotation phases $\phi_0$ and $\phi$,
$$
\frac{1+(p^2-1)\sin^2\theta\cos^2\phi}
{1+(p^2-1)\sin^2\theta\cos^2\phi_0}=q^2,
$$
so, with $0< q\le 1$, that is, $\phi\le\phi_0$ (due to symmetry, we only
need to consider the interval $0\le\phi\le\pi/2$), we define iso-$q$ contours
in the $(\phi,\theta)$ plane (for given $p,\phi_0$) by
$$
r(q,p,\phi_0,\phi):=\frac{q^2-1}{(p^2-1)(\cos^2\phi-
q^2\cos^2\phi_0)},
$$
so, to have viable solutions for $\theta_q$ from $\sin^2\theta_q=r$, we must have $p\le q$, $\phi\le\phi_0$, and 
$$
\cos^2\phi\ge \frac{q^2-1}{p^2-1}+q^2\cos^2\phi_0:=s(q,p,\phi_0)\ge\cos^2\phi_0,
$$
so $\phi$ exist for given $p,q,\phi_0$ only if $s\le 1$; that is,
$$
\cos^2\phi_0\le \frac{p^2-q^2}{q^2(p^2-1)}:=t(q,p).
$$
Denoting
$$
\tilde s(q,p,\phi_0):=\arccos\sqrt{s(q,p,\phi_0)},\quad \tilde t(q,p):=\arccos\sqrt{t(q,p)},
$$
 our CDF is thus
\begin{equation}
\begin{split}
C_q(q)&=\int_0^q f(p)\int_{\tilde t(q,p)}^{\pi/2} \int_0^{\tilde s(q,p,\phi_0)}
\int_{\theta(q,p,\phi_0,\phi)}^{\pi/2}\sin\theta' 
\,d\theta' \,d\phi \,d\phi_0 \,dp\\
&=\int_0^q f(p)\int_{\tilde t(q,p)}^{\pi/2} \int_0^{\tilde s(q,p,\phi_0)}
\sqrt{1-r(q,p,\phi_0,\phi)} \,d\phi \,d\phi_0 \,dp.
\end{split}\label{Cqp}
\end{equation}
Again, we can include the $\beta$-distribution by expanding the integral in the same way as with Eq.\ \eqref{CApx}.

The basis function $G_i(q)$ for a given $p_i$ in the two-point brightness scatter case is, from Eq.\ (\ref{Cqp}),
\begin{equation}
G_i(q)=\left\{\begin{array}{rl}
0, & q\le p_i\\
\int_{\tilde t(q,p_i)}^{\pi/2} \int_0^{\tilde s(q,p_i,\phi_0)}
\sqrt{1-r(q,p_i,\phi_0,\phi)} \,d\phi \,d\phi_0 ,& q>p_i.
\end{array}\right.
\end{equation}
Although the $\phi$-integral can be given in terms of elliptic functions, this is best computed by evaluating 
the double integral numerically. The maximum value of $G_i(q)$ is obtained at $q=1$:
$$
G_i(1)=\int_0^{\pi/2}\int_0^{\phi_0}\,d\phi\,d\phi_0=\frac{\pi^2}{8}.
$$
Our basis functions $G_i$ are closed-form expressions of those computed by Monte-Carlo sampling in \citet{szabo}.
These can be used to determine the $p$-distribution, although we found the accuracy inferior to the solution based on the variation observable $\eta$, which was to expected.

In principle, we can expand $G_i$ to $G_{ij}(q)$ for a $(p_i,\beta_j)$-grid in the same way that $F_i$ were expanded to $F_{ij}$.
However, a notable difference between the two-index basis functions of $A$- or $q$- data is that the $G_{ij}(q)$ all reach their maxima at the same point $q=1$ since the two-point comparison can always contain two equal brightnesses for any $p$ and $\beta$. Thus the $\beta_j$-curves of the $G_{ij}(q)$ of a given $p_i$ form a curve family with the same abscissae for the minimum ($q=p_i$) and maximum ($q=1$); that is, members of the family can easily be mimicked by a superposition of other members unlike in the case of $F_{ij}(A)$. A number of simulations indeed confirmed that  $G_{ij}$ are not usable for solving the inverse problem in practice; that is, $\beta$-information is not recoverable from $q$-data. Adding prior assumptions on the joint distribution did not help either, as it resulted in too heavy regularization,
causing the solution to become almost entirely prior-based.

The same problem plagues the lightcurve slope estimate of \citet{mcneill} from two neighbouring points, exacerbated by the effect of the rotation period $P$. For any basis function, the abscissae for the two-point ratio $s$ lie at some minimum $s_m(p,P)$ and maximum $s=1$. The same abscissae apply not only to all $\beta$, but also to infinitely many other combinations of $p$ and $P$ that yield the same $s_m$. Thus any basis function can be mimicked by numerous different superpositions of basis functions at other $p$, $P$, and $\beta$, making the solution of the inverse problem ambiguous without heavy prior assumptions.

\section{Stability properties of shape distribution from observed brightness variation} \label{app:stability}

We can analyse the inverse problem of determining $f(p)$ with the same approach as in lightcurve inversion \citet{genproj}: we expand both the observed $C(A)$ and $f(p)$ as function series, and examine the relationship between their coefficients. This shows if all coefficients of $f(p)$ can be determined, and also how fast their errors grow as a function of their degree.

Let us expand $f(p)$ as the polynomial
$$
f(p)=\sum_{n=1}^\infty c_n p^n;\quad p\in[0,1].
$$
For isotropic $\theta$,
$$
C(A)=\sum_n c_n \int_0^A p^n\frac{\sqrt{A^2-p^2}}{\sqrt{1-p^2}}\, dp,
$$
and from tables of integrals we find that this is
$$
C(A)=A^2\sum_n c_n A^n \frac{1}{n+1} F_1(\frac{n+1}{2};\frac{1}{2},-\frac{1}{2};\frac{n+1}{2}+1;A^2,1),
$$
where $F_1$ is the Appell hypergeometric function. This form can be transformed into the usual Gauss hypergeometric function $_2F_1$ so that
$$
C(A)=A^2\sum_n c_n A^n k_n G_n(A),
$$
where
$$
G_n(x)=\ _2F_1(\frac{n+1}{2},-\frac{1}{2};\frac{n+4}{2};x^2)=\sum_j^\infty b_j^n x^{2j},
$$
with
$$
b_j^n=\frac{(\frac{n+1}{2})_j(-\frac{1}{2})_j}{j!(\frac{n+4}{2})_j}; \quad (a)_j=\frac{\Gamma(a+j)}{\Gamma(a)}
$$
(so $(a)_0=1=b_0^n$), and
$$
k_n=\frac{\sqrt{\pi}}{2(n+1)}\frac{\Gamma(\frac{n+3}{2})}{\Gamma(\frac{n+4}{2})},
$$
so $k_n\ne 0$ decreases monotonously as $n$ increases, and $\lim_{n\rightarrow\infty} k_n=0$. The decrease is moderate, approximated by, for example, $\sim (n+1)^{-1}[\log (n/2+3)]^{-3/2}$ for $n<100$.
For the gamma function, $\Gamma(n+1/2)=\sqrt{\pi}(2n-1)!!/2^n$ and $\Gamma(n)=(n-1)!$.

Suppose the observed $C(A)$ is expanded (to hold for $0\le A\le 1$) as
$$
C(A)=A^2\sum_{n=1}^\infty a_n A^n.
$$
Then
\[
\begin{split}
&a_1=c_1k_1\Rightarrow c_1=a_1/k_1;\quad c_2=a_2/k_2;\\
&a3=c_3k_3+c_1k_1b_1^1\Rightarrow c_3=(a_3-c_1k_1b_1^1)/k_3,
\end{split}
\]
and so on recursively; that is,
$$
c_n=\frac{1}{k_n}(a_n-\sum_{i=1}^{[n]-1} c_{n-2i} k_{n-2i} b^{n-2i}_i),
$$
where $[n]$ is $(n+1)/2$ or $n/2$ for, respectively, odd or even $n$.
Thus, all coefficients $c_n$ are obtained, and their error grows as $1/k_n$, which is much slower than in, for example, lightcurve inversion.

We note that we can write a formal, more user-friendly one-to-one mapping between the polynomial coefficients
determining $f(p)$ and $C(A)$.
If $p\in[0,1[$, and we expand (assuming $f(p)$ to vanish fast enough when $p\rightarrow 1$)
$$
\frac{f(p)}{\sqrt{1-p^2}}=\sum_{n=1}^\infty d_n p^n,
$$
we have
\[
\begin{split}
C(A)&=\sum_n d_n \int_0^A p^n \sqrt{A^2-p^2} \,dp\\
&=A^2\sum_n d_n A^n\frac{_2F_1(-\frac{1}{2},\frac{n+1}{2};\frac{n+3}{2};1)}{n+1},
\end{split}
\]
which is simply
$$
C(A)=A^2\sum_n d_n k_n A^n,
$$
so we obtain a simple relationship between data and model:
$$
d_n=\frac{a_n}{k_n}.
$$

We also note that the above applies to the general triaxial ellipsoid as well.  
Let us now have a fixed $c\ne1$, $b=1$, and $a=1/p$. Then
$$
A^2=\frac{p^2\sin^2\theta+c^{-2}\cos^2\theta}{\sin^2\theta+c^{-2}\cos^2\theta},
$$
so the iso-$A$ curves are given by
$$
\cos^2\theta_{A3}(p):=g_{A3}(p)=\frac{A^2-p^2}{h(A)-p^2},
$$
where
$$
h(A):=A^2(1-c^{-2})+c^{-2}.
$$
Now
$$
C(A)=\sum_n c_n \int_0^A p^n\sqrt{g_{A3}(p)}\, dp,
$$
and this is
$$
C(A)=\frac{A^2}{\sqrt{h(A)}}\sum_n c_n A^n \frac{1}{n+1} F_1(\frac{n+1}{2};\frac{1}{2},-\frac{1}{2};\frac{n+1}{2}+1;\frac{A^2}{h(A)},1).
$$
This can be used to define a series expansion for the observed $C(A)$ with new basis functions instead of polynomials, so we have the same kind of one-to-one correspondence as above.

\end{appendix}


\begin{thebibliography}{}

% Form:
% \bibitem[authors(year)]{label}
% Lastname, F., ..., paper, vol, starting_page
% (n.b. no article title!)

\bibitem[Ali-Lagoa et al.(2014)]{Ali-Lagoa_et_al_2014}
Ali-Lagoa, V., Lionni, L., Delbo, M et al. 2014, Astron. \& Astrophys., 561, A45

\bibitem[Bowell et. al.(2014)]{Bowell_et_al_2014}
Bowell, E., Oszkiewicz, D.A., Wasserman, L.H. et al. 2014,
Meteoritics and Planetary Science, 49, 95

\bibitem[Cibulkov{\'a} et al.(2016)]{cib}
Cibulkov{\'a}, H., \v{D}urech, J., Vokrouhlicky, D., Kaasalainen M., and Oszkiewicz, D. 2016, 
A \&A, 596, A57

\bibitem[Connelly \& Ostro(1984)]{con}
Connelly, R. \& Ostro, S. 1984,
%\emph{Ellipsoids and lightcurves}
Geometriae Dedicata, 17, 87%--98.

\bibitem[\v{D}urech et al.(2009)]{sparseD}
\v{D}urech, J., Kaasalainen, M., Warner, B., et al. 2009,
A \&A, 493, 291

\bibitem[\v{D}urech et al.(2015)]{aiv}
\v{D}urech, J., Carry, B., Delbo, M., Kaasalainen M., and Viikinkoski, M. 2015, Asteroid Models From Multiple Data Sources,
in Asteroids IV, ed. P. Michel et al., (U. Arizona, Tucson), 183

\bibitem[\v{D}urech et al.(2016)]{wise}
\v{D}urech, J., Hanu\v{s}, J., Ali-Lagoa, V., Delbo, M., and Oszkiewicz, D. 2016. WISE data and sparse photometry used for shape reconstruction of asteroids, in Asteroids: New Observations, New Models, Proceedings of the International Astronomical Union, IAU Symposium, Volume 318, 170

\bibitem[{\v D}urech et al.(2016b)]{Durech_et_al_2016} {\v D}urech, J., Hanu{\v s}, J., Oszkiewicz, D., \& Van{\v c}o, R.\ 2016b, \aap, 587, A48 

\bibitem[Hanu{\v s} et al.(2011)]{Hanus_et_al_2011} Hanu{\v s}, J., {\v D}urech, J., Bro{\v z}, M., et al.\ 2011, \aap, 530, A134 

\bibitem[Hanu{\v s} et al.(2013)]{Hanus_et_al_2013} Hanu{\v s}, J., Bro{\v z}, M., {\v D}urech, J., et al.\ 2013, \aap, 559, A134 

\bibitem[Kaasalainen(2004)]{sparse}
Kaasalainen, M. 2004,
A\&A, 422, L39

\bibitem[Kaasalainen \& Lamberg(2006)]{genproj}
Kaasalainen, M. \& Lamberg, L. 2006,
%\emph{Inverse problems of generalized projection operators}\\
Inverse Problems. 22, 749%--769.

\bibitem[Lagerkvist et al.(1987)]{uapc01}
Lagerkvist, C., Barucci, M.~A, Capria, M.~T., et al.\ 1987,
Asteroid photometric catalogue., ed. C.-I. Lagerkvist, M. A. Barucci, M. T. Capria, M. Fulchignoni, L. Guerriero, E. Perozzi, \& V. Zappala

\bibitem[Mainzer et al.(2011)]{Mainzer_et_al_2011_Preliminary_results_from_NEOWISE}
Mainzer, A., Bauer, J., Grav, T. et al. 2011,
Astrophys. J., 731, 53

\bibitem[McNeill et al.(2016)]{mcneill}
McNeill, A., Fitzsimmons, A., Jedicke, R. et al. 2016,
%\emph{Brightness variation distributions among main belt asteroids from sparse light curve sampling with Pan-STARRS}
MNRAS 459, 2964%--2972.

\bibitem[Oszkiewicz et al.(2011)]{Oszkiewicz_et_al_2011}
Oszkiewicz, D.A., Muinonen, K., Bowell, E. et al. 2011,
AAPP, 89, C1V89S1P072

\bibitem[Piironen et al.(2001)]{uapc02}
Piironen, J., Lagerkvist, C., Torppa, J., Kaasalainen, M., \& Warner, B. 2001, in BAAS, 33, 1562

\bibitem[Szabo \& Kiss(2008)]{szabo} Szab\'o, G. \& Kiss, L. 2008,
%\emph{The shape distribution of asteroid families: Evidence for evolution driven by small impacts}\\
%ScienceDirect,
Icarus, 196, 135%--143.

\bibitem[Warner et al.(2011)]{Warner_et_al_2011_Save_the_lightcurves}
Warner, B.D., Stephens, R.D., Harris, A.W. 2011,
Minor Planet Bulletin, 38, 172

\end{thebibliography}
\end{document}